\documentclass[fleqn,usenatbib]{mnras}

\usepackage[T1]{fontenc}
\usepackage{ae,aecompl}

\usepackage{graphicx}	% Including figure files
\usepackage{amsmath}	% Advanced maths commands
\usepackage{amssymb}	% Extra maths symbols
\usepackage{float}
\usepackage{euscript}
\usepackage{nicefrac}

\newcommand{\msun}{$\rm M_\odot$}
\title[GCs as mass tracers]{Globular clusters as tracers of the dark matter  content of dwarfs in galaxy clusters}
\author[J. E. Doppel et al.]{
\parbox[t]{\textwidth}{
Jessica E. Doppel$^{1}$\thanks{E-mail: jdopp001@ucr.edu (UCR)}\href{https://orcid.org/0000-0001-5354-4229}{\includegraphics[scale=0.8]{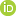}},
Laura V. Sales$^{1}$\href{https://orcid.org/0000-0002-3790-720X}{\includegraphics[scale=0.8]{figs/orcid.png}},
Julio F. Navarro$^{2}$\href{https://orcid.org/0000-0003-3862-5076}{\includegraphics[scale=0.8]{figs/orcid.png}},
Mario G. Abadi$^{3,4}$\href{https://orcid.org/0000-0003-3055-6678}{\includegraphics[scale=0.8]{figs/orcid.png}},
Eric W. Peng$^{5,6}$\href{https://orcid.org/0000-0002-2073-2781}{\includegraphics[scale=0.8]{figs/orcid.png}},
Elisa Toloba$^{7}$\href{https://orcid.org/0000-0001-6443-5570}{\includegraphics[scale=0.8]{figs/orcid.png}} and
Felipe Ramos-Almendares$^{3,4}$
}
\\
\\
$^{1}$University of California Riverside, 900 University Ave., Riverside CA 92521, USA\\
$^{2}$Department of Physics and Astronomy, University of Victoria, Victoria, BC V8P 5C2, Canada\\
$^{3}$CONICET-Universidad Nacional de C\'ordoba, Instituto de Astronom\'ia Te\'orica y Experimental (IATE), C\'ordoba, Argentina\\
$^{4}$Observatorio Astron\'omico, Universidad Nacional de C\'ordoba,  C\'ordoba, Argentina\\
$^{5}$Department of Astronomy, Peking University, 5 Yiheyuan Road, Beijing, China 100871\\
$^{6}$Kavli Institute for Astronomy and Astrophysics, Peking University, 5 Yiheyuan Road, Beijing, China 100871\\
$^{7}$Department of Physics, University of the Pacific, 3601 Pacific Avenue, Stockton, CA 95211, USA\\
}

\date{Accepted XXX. Received YYY; in original form ZZZ}

\pubyear{2015}

\begin{document}
\label{firstpage}
\pagerange{\pageref{firstpage}--\pageref{lastpage}}
\maketitle

% Abstract of the paper
\begin{abstract}
Globular clusters (GCs) are often used to estimate the dark matter content of galaxies, especially dwarfs, where other kinematic tracers are lacking. These estimates typically assume spherical symmetry and dynamical equilibrium, assumptions that may not hold for the sparse GC population of dwarfs in galaxy clusters. We use a catalog of GCs tagged onto the Illustris simulation to study the accuracy of GC-based mass estimates. We focus on galaxies in the stellar mass range 10$^{8} - 10^{11.8}$ M$_{\odot}$ identified in $9$ simulated Virgo-like clusters. Our results indicate that mass estimates are, on average, quite accurate in systems with GC numbers $N_{\rm GC} \geq 10$ and where the uncertainty of individual GC line-of-sight velocities is smaller than the inferred velocity dispersion, $\sigma_{\rm GC}$. In cases where $N_{\rm GC} \leq 10$, however, biases may result depending on how $\sigma_{\rm GC}$ is computed. We provide calibrations that may help alleviate these biases in methods widely used in the literature. As an application, we find a number of dwarfs with $M_{*} \sim 10^{8.5}\, M_{\odot}$ (comparable to the ultradiffuse galaxy DF2, notable for the low $\sigma_{GC}$ of its $10$ GCs) with $\sigma_{\rm GC} \sim 7$ - $15\; \rm km \rm s^{-1}$. These DF2 analogs correspond to relatively massive systems at their infall time ($M_{200} \sim 1$ - $3 \times 10^{11}$ $M_{\odot}$) which have retained only $3$-$17$ GCs and have been stripped of more than 95$\%$ of their dark matter. Our results suggest that extreme tidal mass loss in otherwise normal dwarf galaxies may be a possible formation channel for ultradiffuse objects like DF2.
\end{abstract}

%%%%%%%%%%%%%%%%%%%%%%%%%%%%%%%%%%%%%%%%%%%%%%%%%%

%%%%%%%%%%%%%%%%% BODY OF PAPER %%%%%%%%%%%%%%%%%%

\section{Introduction}

%some words to make the references appear 
%
Pioneering models of galaxy formation established that dwarf galaxies must have been inefficient at forming stars in order to reconcile the observed abundance of faint galaxies with the number of dark matter halos predicted in hierarchical formation models like the Cold Dark Matter scenario \citep[CDM, ][]{White1978,White1991}. These ideas were confirmed by studies of rotation curves in late-type dwarf irregulars \citep{Carignan1988,Broeils1992,Cote2000,Swaters2009}, and of  the stellar kinematics of stars in dwarf spheroidals of the Milky Way and Local Group \citep{Walker2007,Simon2007,Strigari2008,Kirby2014}, which demonstrated that dwarf galaxies are indeed heavily dark matter dominated.

The precise distribution of the dark matter compared to the luminous mass in these systems is less well known. Although dark-matter-only simulations suggest a universal mass profile \citep{Navarro1996}, observations reveal instead a rich diversity of mass profiles in the inner few kiloparsecs of gas rich dwarfs \citep{Oman2015}. This diversity problem has elicited a number of proposals, which can be grouped into three main categories: (i) baryonic and feedback processes that may alter the dark mass profile \citep{Navarro_cores_1996,Pontzen2012,Read2016}; (ii) uncertainties in the interpretation of rotation curves due to non-circular motions and/or triaxiality \citep{Hayashi2006,Pineda2017,Read2016b,Oman2019}; and (iii) more radical changes to the nature of dark matter, such as light axions \citep{Marsh2015}, or the inclusion of a non-negligible self-interaction cross section \citep{Spergel2000,Firmani2000,Creasey2017,Santos2020}. Additionally, some baryon-only dwarfs could be the result of more exotic processes, such as energetic AGN outflows generating gas shells that fragment into individual dwarf-like mass objects (e.g. \citet{Natarajan1998}).

Early-type (i.e., spheroidal) dwarfs may provide important and independent constraints on these ideas. Common in high-density environments, such as groups and clusters, or simply as satellites of MW-like hosts, early type dwarfs are gas-poor, dispersion-dominated systems whose dark matter content may shed light on our understanding of dark matter and its interplay with baryons during galaxy assembly.

The lack of gas means that studies of early-type dwarfs require a different dynamical tracer. The relative brightness and extended spatial distributions of globular clusters (GCs) make them competitive kinematic tracers of galaxy mass. Indeed, in elliptical galaxies, GC studies have enabled constraints on enclosed mass and dark matter fractions with accuracy comparable to studies of HI rotation curves in spirals \citep{Alabi2016,Alabi2017,Longobardi2018}.

Extending these studies to dwarf galaxies is challenging because the number of bright GCs in dwarfs is substantially smaller than in massive systems. For example, several hundred GCs have been used to map the mass distribution around bright galaxies like M87 in Virgo \citep{zhu2014, li2020} and several dozens for luminous ellipticals in the SLUGGS survey \citep{Forbes2017c}. For comparison, in dwarfs with $M_* \leq 10^9$\msun, this quickly reduces to fewer than $\sim 20$ GCs per galaxy. 

Despite this, GC studies have already yielded important constraints on the dark matter content of dwarf ellipticals (dE) in the Virgo cluster \citep{Toloba2016} and, more recently, on ``ultra-diffuse'' galaxies \citep[UDGs, ][]{vanDokkum2016}, where kinematic measurements of the unresolved stellar population are hindered by their low surface brightness \citep{Beasley2016a,Toloba2018,vanDokkum2018a}

As in late-type dwarfs, GC studies of early-type dwarfs also suggest a wide range of dark matter content, with important consequences for the formation paths of UDGs and, potentially, for the nature of dark matter. Of particular interest is the discovery of at least one UDG dwarf, NGC 1052-DF2 (hereafter ``DF2'', for short) , where the extremely low value of the velocity dispersion of the GC \citep{vanDokkum2018a,Wasserman2018} and stellar \citep{Danieli2019} populations hint at little to no dark matter content for this dwarf with estimated stellar mass $M_* \sim 3 \times 10^8$\msun. Although the exact value of the velocity dispersion of GCs (as well as the distances to galaxy \citep{Trujillo2019} is still  being debated ($\sigma_{\rm GC} \sim 5$-$10$ km/s) and may depend on model assumptions \citep{vanDokkum2018b,Martin2018,Laporte2019}, it is at least a factor of $\sim 3$ smaller than that measured for the similar UDG system DF44, which has comparable stellar mass \citep[$\sigma_{\rm GC} \sim 35$ km/s,][]{vanDokkum2019c}, which is broadly consistent with GC velocity dispersions of other dEs of similar stellar mass in the Virgo cluster \citet{Toloba2016}.

Another puzzling dwarf also associated to NGC 1052 is DF4, a UDG where the measured GC velocity dispersion $\sigma_* \sim 4.2$ km/s leaves little room for dark matter \citep{vanDokkum2019a}, though the distance to this system, as with that of DF2, is still under discussion \citep{Monelli2019}. The existence of dwarfs with similar stellar mass but such a wide range of morphology and dark matter content presents a clear challenge to current galaxy formation models. 

Several scenarios have been proposed to form UDGs, including (i) feedback effects combined with environmental gas removal \citep{Chan2018,DiCintio2017,Jiang2019,Tremmel2019}; (ii) unusually large dark matter halos or failed MW-like galaxies \citep{vanDokkum2015}; (iii) dwarf halos with large spin \citep{Amorisco2016, mancerapina2020}; (iv) puffed up stellar systems due to the removal of gas to ram-pressure stripping \citep{Safarzadeh2017}; (v) tidal stripping of cored dark matter halos \citep{Carleton2019}; or (vi) a mixed population made of both: born low-surface brightness dwarfs and tidal remnants of cuspy halos from more massive tidally stripped galaxies \citep{Sales2020}.

%Several scenarios have been proposed to form UDG galaxies, including (i) feedback effects combined with environmental gas removal \citep{Chan2018,DiCintio2017,Jiang2019,Tremmel2019}; (ii) unusually large dark matter halos or failed MW-like galaxies \citep{vanDokkum2015}; (iii) dwarf halos with large spin \citep{Amorisco2016} or unusually large/early mergers \citep{Wright2020}; (iv) puffed up stellar systems due to the removal of gas to ram-pressure stripping \citep{Safarzadeh2017} or tidal heating \citep{Carleton2020}: (v) tidal stripping of cored dark matter halos \citep{Carleton2019}; or (vi) a mixed population made of both: born low-surface brightness dwarfs and tidal remnants of cuspy halos from more massive tidally stripped galaxies \citep{Sales2020}.

Encouragingly, the observational evidence seems to support a variety of formation paths for UDGs. For instance, the number of associated GCs varies widely, from $\sim 30$ in DF17 \citep{Peng2016} to some UDGs in Coma with no associated GCs at all \citep{Beasley2016b,Lim2018}. A systematic study of UDGs and their GCs in the Virgo cluster also confirms the trends found in Coma \citep{Lim2020}. More detailed, kinematical studies of 3 UDGs in Virgo have also revealed wide variations in enclosed dark matter mass, including one object, VLSB-D, with clear signatures of ongoing tidal disruption \citep{Toloba2018}. Intriguingly, of the 3 UDGs studied, VLSB-D has the largest stellar mass ($M_* \sim 7.9 \times 10^8$\msun) but also the lowest GC velocity dispersion, $\sigma = 16^{+6}_{-4}$ km/s.

It is tempting then to consider the following hypothesis: could the low velocity dispersion measured for some UDGs (DF2, DF4, VLSB-D) be explained as a result of tidal effects in cluster or group environments? Or, in other words, can simulations reproduce a GC velocity dispersion as low as $\sigma_{\rm GC} \sim 10$ km/s (or lower) in a galaxy with stellar mass as high as $M_*\sim 3 \times 10^8\, M_\odot$?
Analytical arguments, combined with the cosmological hydrodynamical simulations presented in \citet{Sales2020}, seem to suggest that this is indeed possible, but more detailed work is needed to fully validate this possibility.

We study these issues here using a catalog of GCs tagged onto the Illustris simulation \citep{ramalm2020}. The simulation follows the dynamical evolution of dwarfs in clusters, providing an ideal tool to quantify the effects of tidal disruption, departures from equilibrium, and scarcity of tracers. We further use the simulations to look into the tidal disruption formation scenario for objects like DF2. Our paper is organized as follows. The GC model and galaxy selection criteria are described in Sec.~\ref{sec:gcs} and \ref{sec:sample}. We evaluate the accuracy of mass estimators in Sec.  ~\ref{sec:mass}, with special emphasis on different methods to measure velocity dispersion, the number of targets, and the effects of tidal disruption. In Sec.~\ref{sec:df2} we use our simulated galaxies and GCs to look for DF2 analogs. We conclude and summarize our main results in Sec.~\ref{sec:concl}.

\section{Methods}
\label{sec:gcs}

We use the highest resolution run of the Illustris cosmological, hydrodynamical simulation (Illustris-1) \citep{vogelsberger2014a, Vogelsberger2014b, genel2014, sijacki2015}. The simulation has a box size of 106.5 Mpc on a side and assumes cosmological parameters consistent with the  WMAP9 \citep{hinshaw2013} results.
At the resolution used here, the mass per particle is  $1.3 \times 10^6$ M$_{\odot}$ and $6.26 \times 10^6$ M$_{\odot}$ for the baryonic and dark matter components respectively, with a maximum gravitational softening length of 710 pc. 

The galaxy formation model used by Illustris includes stellar evolution and supernova feedback, black hole growth and mergers, AGN feedback, as well primordial and metal line cooling, among others. The simulation matches a number of observables well, including the Tully-Fisher relation \citep{Torrey2014, Vogelsberger2014b}, the cosmic star formation density \citep{genel2014}, the galaxy mass and luminosity functions \citep{Vogelsberger2014b}, and the wide range of colors and morphologies of the present-day galaxy population \citep{sales2015, snyder2015, rodriguez-gomez2017}.

\subsection{Galaxy sample}
Our galaxy sample consists of members of the $9$ most massive clusters, with masses comparable to the Virgo cluster ($M_{200} > 8\times 10^{13}$\msun), in Illustris-1. Throughout this paper, we shall define virial quantities as measured at the radius containing $200$ times the critical density of the universe. Halos and subhalos are identified using a combination of a friends-of-friends algorithm \citep[FoF, ][]{davis1985} and {\sc subfind} \citep{springel2001,dolag2009}.  We use the {\sc sublink} merger trees \citep{rodriguez-gomez2015} to trace the assembly of these clusters back in time. We follow the infall and posterior evolution of satellite galaxies identified in these $9$ host clusters at $z=0$, focusing on those in the stellar mass range  $10^{8}<M_*< 6 \times 10^{11}$\msun. The minimum stellar mass cut implies a minimum of $\sim 60$ stellar particles in our objects, which we consider to be sufficiently resolved for the purpose of this analysis. Additionally, we require a minimum stellar mass at infall $M_* \geq 5 \times 10^8$\msun which guarantees on average more than 16,000 particles including dark matter, gas, and stars at infall. Our simulated galaxy catalog contains a total of $3777$ satellite galaxies, and it records the {\it infall time} of each galaxy as the last time, before accretion, that it was the central galaxy of its own FoF halo.

%%%%%%%%%%%%%%%%%%%%%%%%%%%%%%%%%%%%%%%%%%%%%%%%%%%
\begin{figure*}
    \centering
    \includegraphics[width=2\columnwidth]{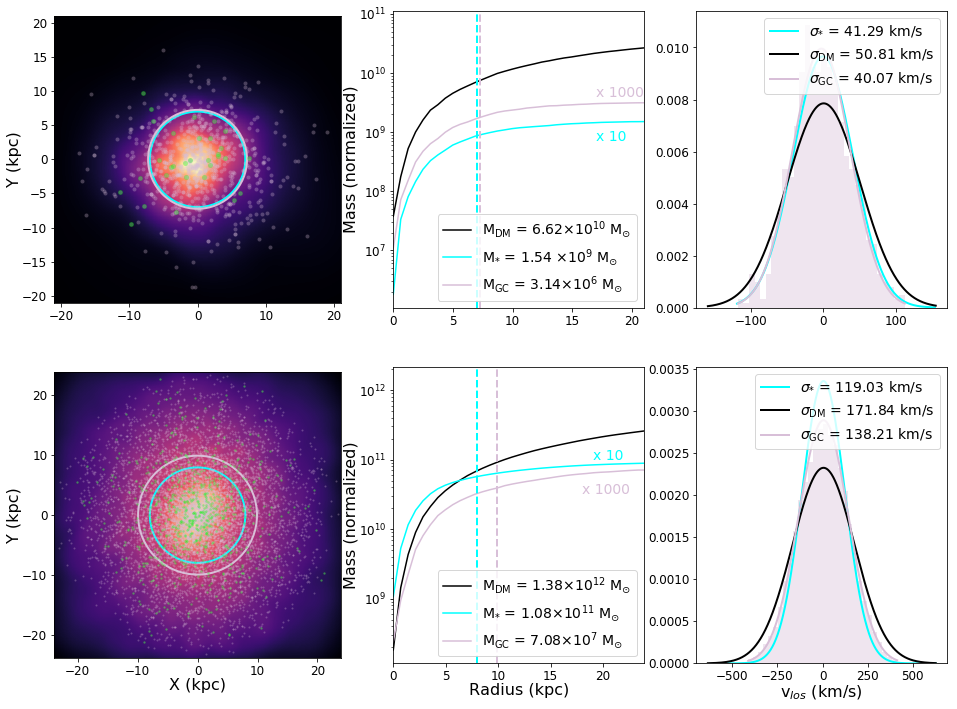}
	\caption{{\it Left:} XY projections of the stellar component (color scale in the background), GC candidates (pinkish points), and realistic GCs (green points), for a dwarf galaxy (top) and a Milky Way mass galaxy (bottom) belonging to the largest simulated galaxy cluster in Illustris. Pink and cyan circles indicate the 3D half mass radius of GC candidates and stars, respectively. {\it Middle:} normalized cumulative mass profiles for the dark matter (black curve), stars (blue curve) and GC candidates (pink curve) associated to these galaxies. The mass profile of the GC candidates has been multiplied by a factor of $\times 1000$, and the stellar profile has been normalized by a factor of $\times 10$. Masses as quoted in the legend. Half mass radius of stars and GCs are highlighted with vertical dashed lines. {\it Right:} line-of-sight velocity distributions for the GC candidates of these two galaxies  (pink shade) along with the best fitting Gaussian in the same color. For comparison, we overplot the best-fitting Gaussians for the velocity distribution of the stars (cyan) and the dark matter in black. Note the similarities of their shapes and dispersion, values quoted for the latter for each galaxy.}
    \label{fig:introfig}
\end{figure*}
%%%%%%%%%%%%%%%%%%%%%%%%%%%%%%%%%%%%%%%%%%%%%%%%%%%

\subsection{Adding GCs to Illustris}
\label{ssec:gcs}

Illustris follows the global star formation properties of galaxies but does not have the resolution to form or follow GCs. In our study, GCs are added to the simulation in post processing by tagging selected dark matter particles in galaxy halos to match, on average, the known properties of the GC population and its dependence on halo mass. The method was introduced in \citet{ramalm2020}, where details may be found. We include a brief description here for completeness.

The tagging process takes place, for each galaxy, at its cluster infall time. At that time, the procedure first identifies dark matter particles satisfying a prescribed density distribution; in particular, a  \citet{hernquist1990} profile with scale radius,  $a_{\rm HQ} = \alpha \, r_{\rm NFW}$, where $r_{\rm NFW}$ is the scale radius of the halo's best-fitting NFW profile \citep{Navarro1996}, and $\alpha$ is a parameter that controls the spatial extent of the GC population. We use here two values of $\alpha=0.5$ and $3$ in order to select candidate tracers of the red and blue GC populations respectively. (Our analysis below, however, does not distinguish between these two populations.) Note that this method by construction selects all particles that are consistent with the energy distribution of GCs, which, in general, is a larger set of particles than the typical number of GCs associated to a galaxy. We therefore must subsample the set of \textit{candidate} GCs to obtain a \textit{realistic} population of GCs. This subsampling is done randomly and assumes that the mass of each GC is $10^5$\msun.

%\textcolor{cyan}{It should be noted that this method produces more GC candidate particles than would realistically be part of a galaxy's GC system. Subsampling is required to acquire a realistic population of GCs.}

For these GC candidates, we assume that the total stellar mass of the GC population, $M_{\rm GC}$, scales with halo virial mass in a manner consistent with the results of \citet{Harris2015}. Note that this relation holds at $z = 0$ while our procedure is applied at infall; thus, some adjustments are necessary, as some GCs may be lost to the cluster due to tidal effects. As shown by \citet{ramalm2020}, a simple relation at infall of the form 
\begin{equation}
  M_{\rm GC} = a M_{200}^b,
\label{EqMGCM200}
\end{equation}
with $a = 2.0\times 10^{-7}$, $3.5 \times 10^{-4}$ and $b = 1.15$, $0.9$ for red and blue GCs respectively, matches the \citet{Harris2015} relation well at $z=0$.
These tagged particles are then used to trace the GC population of cluster galaxies after infall, as well as intracluster GC populations, which is made of all GCs stripped from galaxies after infall. At $z=0$, the remaining candidates are subsampled assuming a fixed mass of $m_{\rm GC} = 10^5$\msun\; per GC to determine a realistic number of GCs.

A specific caveat of this procedure is that we tag and follow only the population of {\it surviving} GCs and we do not account --by design-- for the internal evolution of stellar clusters. Instead, our catalog can be used to study the dynamical process that GCs are subjected to within galaxy clusters after each galaxy, with their corresponding GC system, has been accreted into the cluster host.   

The GC catalog created following this procedure has been shown to reproduce, without further adjustment, some key observational properties, including the large scatter in the specific frequency $S_N$ for dwarf galaxies and the formation of an extended and diffuse population of ``intracluster" GCs \citep{ramalm2020}. In this paper we focus on the GC population around each surviving galaxy in the cluster at $z=0$ in order to check to the accuracy of GC-based estimates of the total dynamical masses of cluster galaxies.

%%%%%%%%%%%%%%%%%%%%%%%%%%%%%%%%%%%%%%%%%%%%%%%%%%%%%
\begin{figure}
    \centering
    \includegraphics[width = \columnwidth]{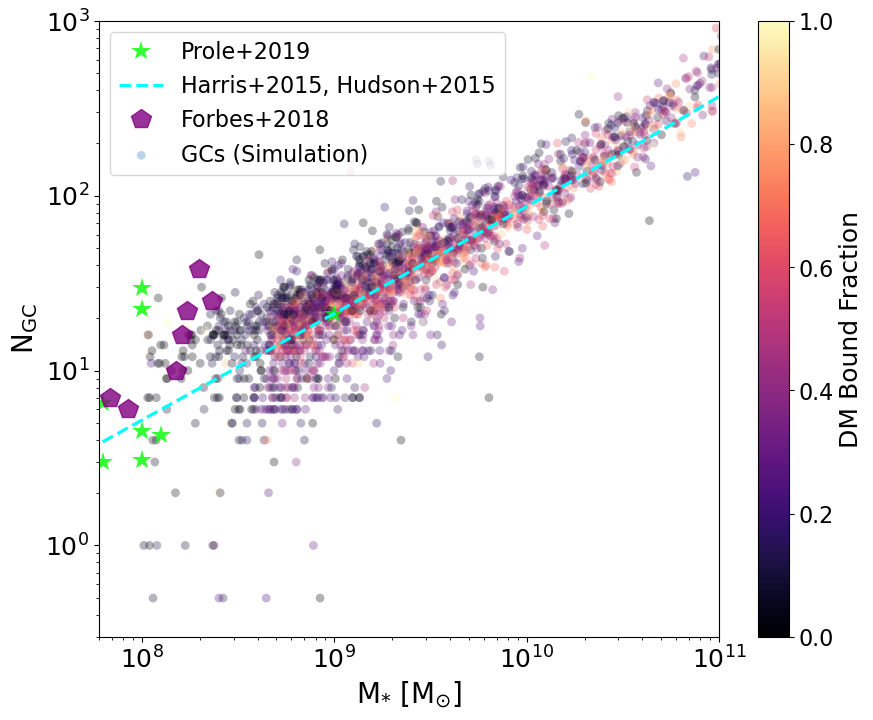}
	\caption{Realistic number of GCs, N$_{\rm GC}$, versus the stellar mass of the host galaxy, M$_{*}$, colored by the bound dark matter fraction (see section ~\ref{sec:sample}). Our GC model is calibrated to reproduce on average the M$_{\rm GC}$ - M$_{\rm halo}$ relation from \citet{Harris2015} (cyan dashed line). Note that tidal stripping partially introduces a significant scatter from galaxy to galaxy, specially on the low mass end. The number of GCs for the lowest mass dwarfs in roughly consistent with observations in \citet{Prole2019} and \citet{Forbes2018} that were not part of the model calibration.}
    \label{fig:ngc}
\end{figure}
%%%%%%%%%%%%%%%%%%%%%%%%%%%%%%%%%%%%%%%%%%%%%%%%%%%%%

%%%%%%%%%%%%%%%%%%%%%%%%%%%%%%%%%%%%%%%%%%%%%%%%%%%%%%%%%%%%%%%%%%%%
\begin{figure}
    \centering
    \includegraphics[width = \columnwidth]{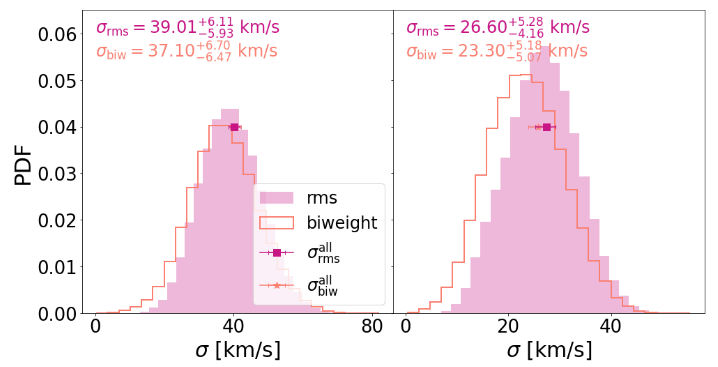}
	\caption{An illustration of the impact of using different definitions for the line-of-sight velocity dispersion $\sigma_{\rm los}$ of GCs in two simulated dwarfs (the left panel corresponds to the dwarf in the top panel of Fig.~\ref{fig:introfig}). Each panel shows the probability distribution function (PDF) of $10^5$ random realizations of $\sigma_{\rm los}$ estimates using subsamplings of $10$ GCs out of $\sim 400$ (left) and $\sim 270$ (right) GC candidate particles for the example dwarfs. We adopt two commonly used definitions, simple r.m.s (filled magenta) and biweight (open orange). These methods can predict slightly differently shaped PDFs, as well as different median values of line-of-sight velocity dispersion as quoted (uncertainties correspond to 25\%-75\% quartiles of the $\sigma$ distributions). The r.m.s and biweight velocity dispersion of the underlying parent sample of $\sim 400$ and $\sim 270$ candidate GCs are shown with squared symbols ($90\%$ confidence interval is also shown as error bars). Although most of the $\sigma$ estimates for each $10$ GC draws would reasonably agree between r.m.s and biweight, for some realizations biweight estimates may underestimate the velocity dispersion compared to its r.m.s definition.}
    \label{fig:pdfs}
\end{figure}
%%%%%%%%%%%%%%%%%%%%%%%%%%%%%%%%%%%%%%%%%%%%%%%%%%%%%%%%%%%%%%%%%%%%
%%%%%%%%%%%%%%%%%%%%%%%%%%%%%%%%%%%%%%%%%%%%%%%%%%%%%%%%%%%%%%%%%%%%
\begin{figure*}
    \centering
    \includegraphics[width = \columnwidth]{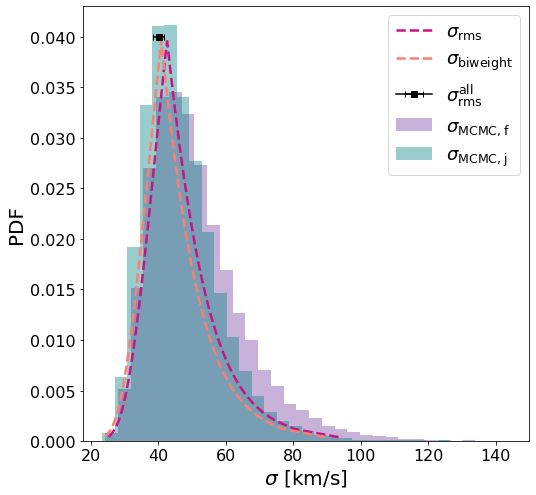}
    \includegraphics[width = \columnwidth]{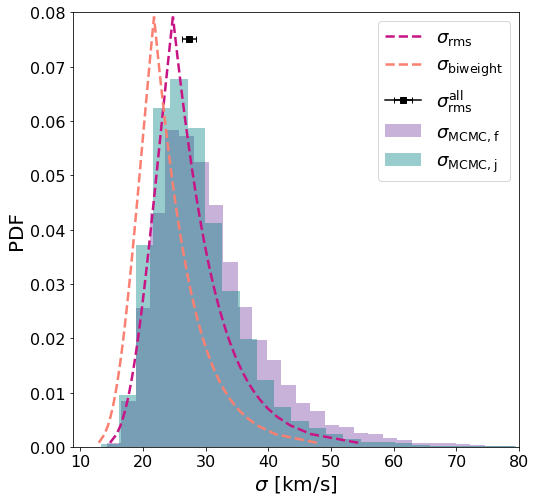}
	\caption{Velocity dispersion and confidence intervals calculated using different methods: r.m.s (magenta), biweight (orange), MCMC with flat prior (purple) and MCMC with Jeffreys prior (teal). We show in each panel one particular realizations of $10$ GCs for the dwarfs in Fig.~\ref{fig:pdfs}. For most possible drawings, the estimates of $\sigma$ with different methods agree well and identify the true underlying (r.m.s) velocity dispersion of the whole $\sim 400$ (left) and $\sim 270$ (right) GC candidate sample ($\sigma_{\rm RMS}^{\rm all}$ square symbol). In some cases as highlighted on the right, biweight may result on a slightly underestimated velocity dispersion compared to the other methods.} 
    \label{fig:pdfs_mcmc}
\end{figure*}
%%%%%%%%%%%%%%%%%%%%%%%%%%%%%%%%%%%%%%%%%%%%%%%%%%%%%%%%%%%%%%%%%%%%

\section{GCs of simulated cluster galaxies}
\label{sec:sample}

We show in Fig.~\ref{fig:introfig} two examples of our simulated galaxies and their GC system. The top and bottom rows correspond to, respectively, a dwarf ($M_* =1.5 \times 10^9$\msun) and a luminous galaxy ($M_* \sim 10^{11}$\msun). The left column shows a stellar map projection on which, to illustrate the tagging procedure,  we superimpose the full population of tagged ``candidate GCs'' (pink points) as well as, in green, the actual particles selected as GCs in this case. The middle column shows the cumulative mass distribution with radius of the stellar component (cyan), dark matter (black) and GC candidates (pink). The GC spatial distribution is similar to that of the stars in the dwarf galaxy, but is significantly  more extended than the stellar component of the more massive galaxy, in good agreement with well-established observational trends \citep[e.g., ][]{georgiev2010,Forbes2017b,Hudson2018,Prole2019}. The line-of-sight velocity distributions of these 3 components are shown in the right column and show that the GC velocity dispersion is comparable to that measured for the stars and the dark matter within three times the half-mass radius of the stars ($r \sim 3 r_{h,*}$). Best-fit Gaussian distributions to each component are also included for comparison.

In order to minimize the number of potential interlopers (i.e., intracluster GCs, or GCs belonging to nearby galaxies) we associate GCs with each individual galaxy using a (3D) distance cut, i.e., $r \leq 3 r_{h,*}$, and a velocity cut,
which applies a
3-$\sigma$ clipping criterion for membership in the line-of-sight velocities (projected in a random direction).
This last step is effective at removing most (although not all) contamination from intracluster GCs and other chance alignments. We have explicitly checked that none of the results presented in this paper change qualitatively if the radial cutoff is varied in the range $2$-$5 r_{h,*}$. GCs satisfying the criteria of distance and velocity are then considered associated to each galaxy and used for dynamical mass estimation.

Fig.~\ref{fig:ngc} shows that our tagging procedure yields realistic numbers of  GCs as a function of their stellar mass. Although by construction the model reproduces the main trend with $M_*$ reported by \citet[][dashed cyan line]{Harris2015} after assuming the $M_*$-$M_{200}$ relation in \citet{Hudson2015}, it is interesting to see the substantial scatter at fixed $M_*$, which results despite the fact that the relation adopted between GC mass and halo mass (Eq.~\ref{EqMGCM200}) is assumed to be scatter-free. 
Moreover, the scatter in number of GCs, $N_{\rm GC}$, increases towards low-mass galaxies, in good agreement with observations \citep{Peng2008,Prole2019, Forbes2018}. For instance, a $M_* \sim 10^9$\msun\; cluster dwarf may show $5$-$20$ GCs, or even none (symbols artificially shifted to $N_{\rm GC}=0.5$). Within the simulation, this scatter results almost exclusively by the effects of tidal stripping in the cluster environment. Indeed, symbols in Fig.~\ref{fig:ngc} are color coded by the DM bound fraction, the ratio of dark matter mass that the {\sc subfind} catalogs records for a galaxy at $z=0$ to that at its infall time. As discussed in \citet{ramalm2020}, tidal stripping effects seem to be critical to explain the origin of the scatter in this relation and of its dependence on mass.

Note that we only tag at infall galaxies with $M_*(t=t_{\rm inf}) \geq 5 \times 10^8$\msun, meaning that all simulated systems in our sample with a present day stellar mass $M_* \leq 5 \times 10^8$\msun result from tidal stripping that has affected its stellar component. This can be seen in the low remaining dark matter bound fraction of most galaxies in that mass range in Fig.~\ref{fig:ngc}. In other words, for $M_*=1$-$5 \times 10^8$\msun\; range at present day, our sample only includes the tidally stripped objects--those that satisfied at infall the tagging criteria with $M_* > 5 \times 10^8$\msun. Simulated dwarfs in this mass range at z = 0 that have never been above the mass threshold for GC tagging are not included in our sample, a topic we return to in Sec.~\ref{sec:df2}.

%population and {\it not} the total population \textcolor{cyan}{of dwarf galaxies within the simulation}, a topic we return to in Sec.~\ref{sec:df2}.

\section{Dynamical mass estimators}
\label{sec:mass}

Under the hypothesis of spherical symmetry and dynamical equilibrium, the mass enclosed by  a collisionless population of tracers within their half-mass radius may be written as:

\begin{equation}
    \rm M(< \rm r_{1/2}) \approx 3\rm G^{-1} \sigma_{los}^2 \rm r_{1/2}
    \label{eq:wolf}
\end{equation}

\noindent
where $\sigma_{\rm los}$ is the line of sight velocity dispersion of the tracers; $r_{1/2}$ is the 3D (de-projected) half-mass radius of the tracers; and $M(< \rm r_{1/2})$ is the total enclosed mass within $r_{1/2}$ ($G$ is Newton's gravitational constant). This mass estimator has been shown to be relatively insensitive to the anisotropy parameter of the orbits (commonly referred to as $\beta$) and to projection effects \citep[see; e.g.,][]{Wolf2010}. Similar formulae have been presented by other groups, but the main variation is in the value of the proportionality constant or in the definition of the radius that the derived enclosed mass applies to. For simplicity, we focus on the reminder of this paper in the \citet{Wolf2010} estimator, but we have explicitly checked that similar conclusions apply when using different models, such as those presented by \citet{Walker2011} or \citet{Errani2018}.
It should be noted that mass estimated derived from the Jeans Equation are sensitive to the assumed underlying mass distribution \citep{Hayashi2018}.

We may use our tagged catalog of GCs to assess how well Eq.~\ref{eq:wolf} recovers the dynamical mass of simulated cluster galaxies in Illustris. One challenge in this case is estimating $\sigma_{\rm los}$, which is well defined when several dozen GCs are present, but is less robust for the small number of tracers available in the regime of dwarf galaxies (see Fig.~\ref{fig:ngc}). In what follows, we will drop the ``line of sight" from the subscript, but we will still refer to the 1D velocity dispersion projected along a random direction, as measured in observations.

\subsection{Velocity dispersion estimates}
\label{ssec:sigma}

%%%%%%%%%%%%%%%%%%%%%%%%%%%%%%%%%%%%%%%%%%%%%%%%%%%%%%%%%%%%%%%%%%%%%%%%%
\begin{figure}
    \centering
    \includegraphics[width = \columnwidth]{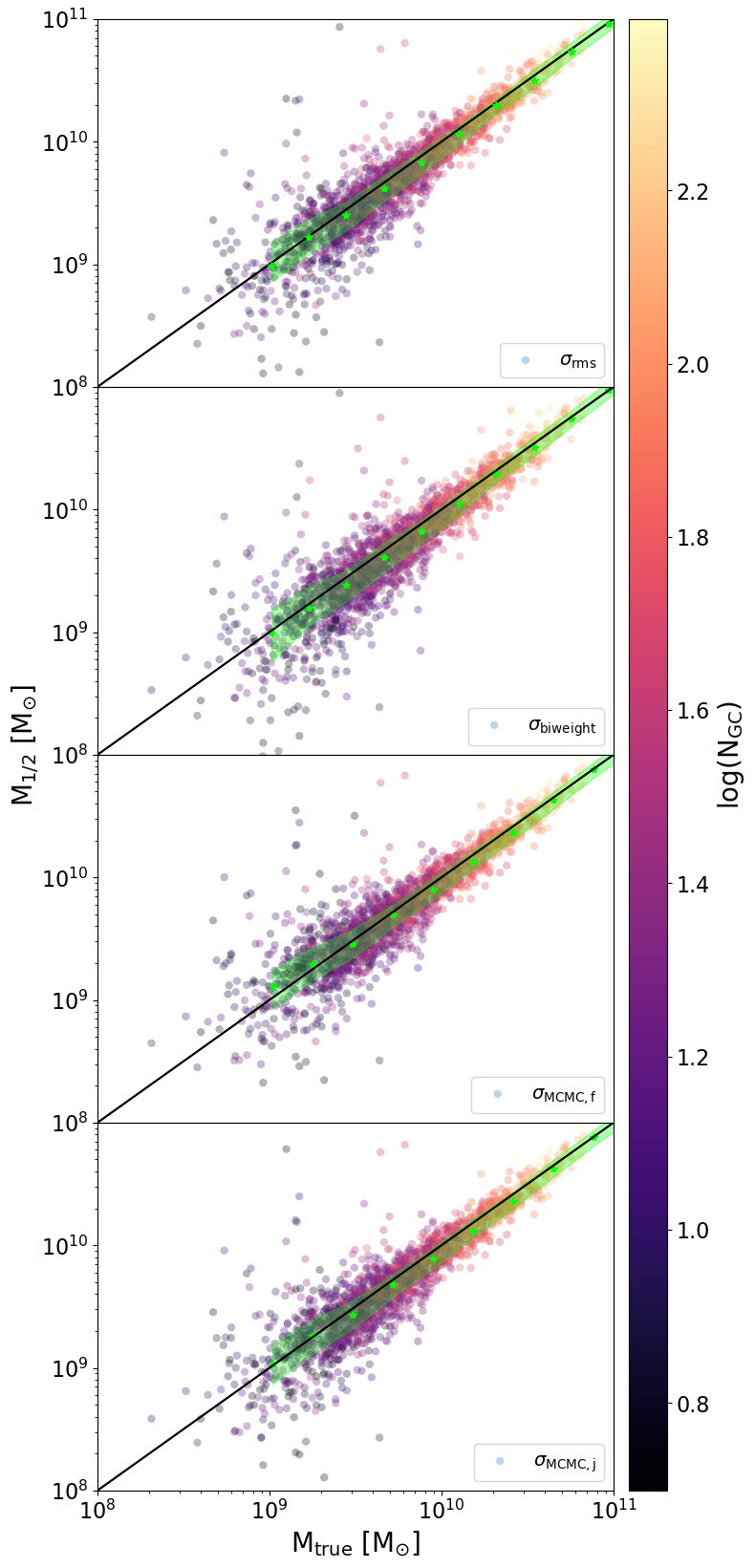}
	\caption{Comparison between the true (x-axis) and estimated (y-axis) dynamical mass measured for simulated galaxies at half-number radius of GCs using \citet{Wolf2010}. Symbols are colored by the log of the number of globular clusters, N$_{\rm GC}$ associated to the host galaxy (color bar). The one-to-one line is shown in black, and the running median of the estimated dynamical mass $M_{1/2}$ at fixed $M_{\rm true}$ is shown in green symbols with $25\%$-$75\%$ quartiles indicated by the green shading. From top to bottom panels correspond to our three $\sigma_{\rm los}$ definitions: rms, biweight and MCMC (flat and Jeffreys priors). On average, all methods to quantify velocity dispersion perform very well to estimate mass on a sufficiently large sample of galaxies. However, the scatter increases for galaxies with a low number of GCs (darker symbols) which might result on significant deviation for individual objects. These deviations from the one-to-one line are systematic depending on the definition of $\sigma_{\rm los}$, as explored in Fig. ~\ref{fig:mederr}.} 
    \label{fig:massest}
\end{figure}
%%%%%%%%%%%%%%%%%%%%%%%%%%%%%%%%%%%%%%%%%%%%%%%%%%%%%%%%%%%%%%%%%%%%%%%%%

Several methods are widely used to compute $\sigma$. 
Here, we consider the following three: (i) the r.m.s of tracer velocities, $\sigma_{\rm rms}$ \citep[see; e.g.,][]{Prada2003}; (ii) the biweight velocity dispersion ($\sigma_{\rm biweight}$) \citep{Beers1990,vanDokkum2018a, Girardi2008, Veljanoski2014}; and (iii) a velocity dispersion, $\sigma_{\rm MCMC}$, estimated using a Markov-Chain-Monte-Carlo (MCMC) method applied to the individual velocities \citep{Hogg2010, Widrow2008, Martin2018, vanDokkum2018b}. Details on each method, as implemented here, are given in Appendix~\ref{app:sigma}.

Each of these methods has their own advantages and disadvantages. The r.m.s velocity dispersion has the advantage of simplicity but it may give biased results in case of non-Gaussian velocity distributions. The biweight method, on the other hand,  is ideal when high levels of contamination are expected since it places more weight towards velocities closer to the median of the distribution, however, it cannot be used for systems with less than 5 tracers \citep{Beers1990}.

The MCMC approach enables a proper treatment of observational uncertainties, but it suffers from sensitivity to the shape of the priors assumed. In this study, we shall compare results using a flat prior distribution or Jeffreys prior, where the latter is usually considered more robust for low number of tracers \citep[e.g., see ][]{Martin2018}. We shall indicate the choice of prior with subscripts ``f" and ``j", respectively, when needed. See Appendix ~\ref{app:sigma} for more details on the prior calculation.

For a given set of tracers, the $\sigma$ probability distribution functions (PDFs) obtained with each of these methods may have slightly different shapes. We show this for the r.m.s and biweight distributions in Fig.~\ref{fig:pdfs} using two dwarfs as examples, the galaxy introduced in the top row of Fig.~\ref{fig:introfig} which is characterized by an intrinsically nearly Gaussian line-of-sight velocity distribution (left panel) and a different dwarf selected to have a non-Gaussian line-of-sight distribution of GC candidates with kurtosis and skewness 0.39 and 1.04, respectively (right panel). The PDFs shown in Fig.~\ref{fig:pdfs} correspond to velocity dispersion estimates obtained from 10$^5$ independent random selections of $10$ GCs from among the $\sim 400$ (left) and $\sim 270$ (right) candidate GC particles that remain associated with these galaxies at $z=0$. 

While the r.m.s (filled magenta) and the biweight (open orange) methods show similar distributions, the biweight shows a systematic (albeit small) trend towards lower $\sigma$ values, especially for non-Gaussian parent samples as illustrated for the dwarf on the right panel. This can be understood in light of the weight assignment for the biweight method, which tends to down-weight values further away from the median of the sample.

Reassuringly, the PDF distribution for these $10$-GC re-sampling shows, in both methods, a well defined peak that agrees well with the velocity dispersion of the underlying parent distribution of $\sim 400$ and $\sim 270$ candidate GC particles (square symbols). However, this exercise highlights one of the main problems with the discreteness of the dynamical tracers: depending on the particular $10$-GC realization, one might obtain estimates far from the true underlying velocity dispersion.

It is interesting to further explore the ability of different methods to estimate the true $\sigma$ under the condition of a limited number of tracers. We do this by selecting one particular realization of $10$ GCs from each of our examples in Fig.~\ref{fig:pdfs}. For each of these two realizations, we estimate the confidence intervals assuming a Gaussian parent distribution in Fig.~\ref{fig:pdfs_mcmc} for the r.m.s (magenta) and biweight (orange) methods. 

We see that in both cases, estimates show a large degree of overlap between r.m.s and biweight, which would be the case for most of the possible $10$-GC re-samplings. However, due to the low-velocity bias seen in biweight in Fig.~\ref{fig:pdfs}, the estimated velocity dispersion with this method may substantially underestimate the true value for some specific samplings (right panel), a possibility that should be kept in mind when working with biweight estimates.

Fig.~\ref{fig:pdfs_mcmc} also shows the corresponding PDF for the MCMC method using both, a flat (purple) and Jeffreys (teal) priors. For each $10$-GC subsampling, the PDF is calculated by a random walk through ($\sigma, \ \langle v \rangle$) parameter space over 10$^5$ iterations using a Gaussian jumping distribution with a dispersion of 5 $\rm km$ $\rm s^{-1}$. For both realizations in Fig.~\ref{fig:pdfs_mcmc} the MCMC method is able to recover the true $\sigma$, with uncertainties that agree well with the simpler r.m.s method.  

Faint dwarf galaxies can have even less than $10$ GCs and the systematic effects explored here for each method may therefore become stronger. In what follows, we use our GC catalog to extend this study to a statistical sample of galaxies in Illustris to explore how the dynamical mass estimates are impacted by the finite number of GCs tracers and underlying assumptions of Gaussianity in the distribution.  
 
\subsection{Mass estimates}
\label{ssec:jeans}

For each of the 3777 simulated cluster galaxies we may use the ``realistic'' number of GCs drawn from the list of candidates to compute the GC half number radius, $r_{1/2}$, and velocity dispersion using different estimators, $\sigma_{\rm rms}$, $\sigma_{\rm biweight}$ and $\sigma_{\rm MCMC}$. We then apply Eq.~\ref{eq:wolf} to estimate their dynamical mass $M_{1/2}$ and compare the results obtained with each estimator with the true mass enclosed within $r_{1/2}$, as measured directly from the particle information in the simulation, $M_{\rm true}$. Fig.~\ref{fig:massest} shows the results, with the solid black line indicating a one-to-one relation and points color-coded by the logarithm of the number of associated GCs used in the calculation. Green symbols show the median in bins of $M_{\rm true}$ and the shading indicates the $25\%$-$75\%$ quartiles.

We find, on average, a remarkably good agreement between the estimated dynamical mass $M_{\rm 1/2}$ and the true mass, supporting the use of simple estimators as that presented in \citet{Wolf2010} to determine the dynamical mass of galaxies using GCs as tracers (similar conclusions hold for estimators proposed in \citet{Walker2011} or \citet{Errani2018}).

This result is not trivial, as many of the assumptions, such as sphericity and/or dynamical equilibrium, on which the estimator is based do not apply to our systems. Our results agree with \citet{laporte2013}, who reported a similar conclusion although applied to {\it stellar} (not GC) tracers in dwarf spheroidal galaxies \textcolor{cyan}{of} the MW. The authors generalized the \citet{bullock2005} method to cosmological triaxial systems \citep[][]{Laporte2013b} and find that the deviations from sphericity are compensated by a trade-off between the changes on the line of sight velocity dispersion and those in the half mass radius that are measured in different projections, canceling out in combination any systematic effect in spherical mass estimators such as Eq.~\ref{eq:wolf}. 

A closer inspection of Fig.~\ref{fig:massest} reveals that systems with a low number of GCs (dark symbols) tend to have larger scatter around the one-to-one line. This coincides with the low mass regime, where dwarf galaxies often have only a few, or up to a dozen GCs. Mass estimators tend to perform poorly with a low number of tracers, specially due to the errors in estimating velocity dispersion and half mass/number radius using only a handful of tracers. 

%%%%%%%%%%%%%%%%%%%%%%%%%%%%%%%%%%%%%%%%%%%%%%%%%%%%%%%%%%%%%%%%%%%%
\begin{figure}
    \centering
    \includegraphics[width=\columnwidth]{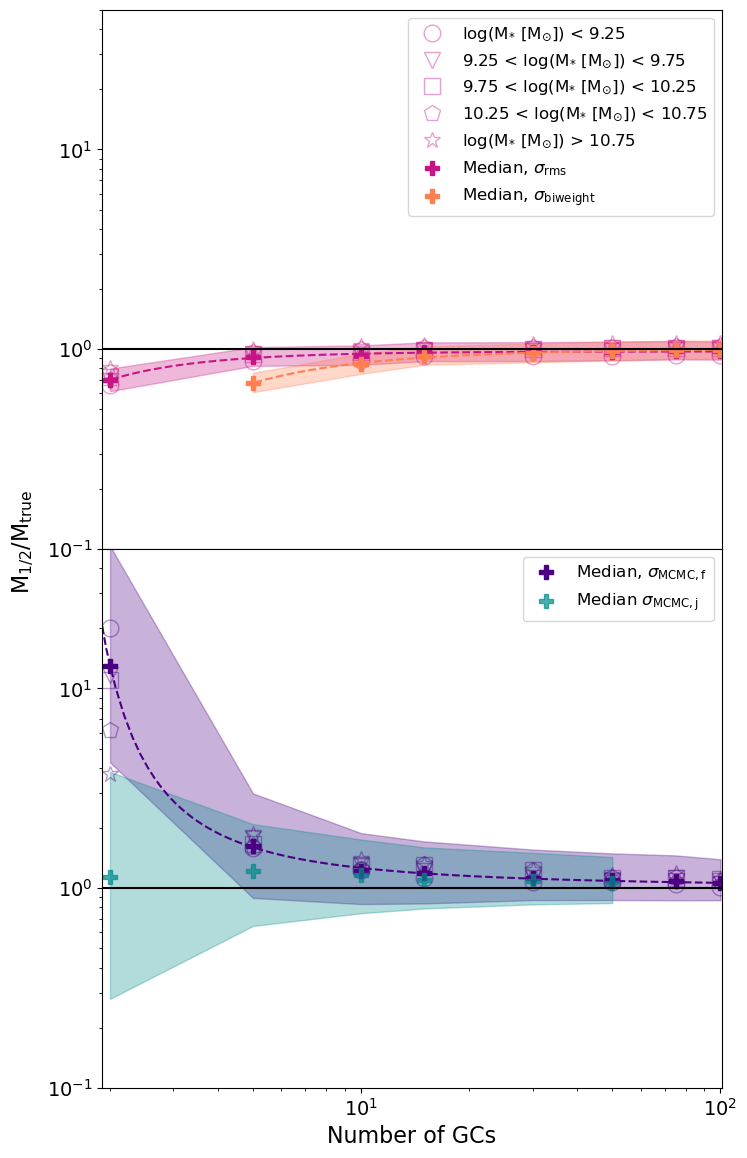}
	\caption{Median of the ratio of the estimated mass to the true mass, $M_{1/2}$/$M_{\rm true}$ as a function of number of GCs used in the estimate of $\sigma_{\rm los}$ following: the rms (magenta) and biweight (orange) in the top panel and MCMC  methods with flat (purple) and Jeffreys (teal) priors in the bottom. Filled symbols show the median, shaded area the quartiles. We find no significant trend with stellar mass of the galaxies once N$_{\rm GC}$ is fixed (see open symbols). However, we find a strong trend with the number of tracers: $\sigma_{\rm rms}$ and $\sigma_{\rm biweight}$ tend to underpredict the dynamical mass while $\sigma_{\rm MCMC}$ overpredict the mass for a low number of GCs. These systematic trends can be corrected using a simple calibration (see dashed lines) shown in Eq.~\ref{eq:errmcmc} and ~\ref{eq:err_rms} with coefficients listed in Table~\ref{tab:err_params}. Note that mass estimates are accurate for galaxies with a sufficiently large number of tracers, for example $M_{1/2}$ is within 10-15\% from the true mass for galaxies with N$_{\rm GC} \geq 30$.} 
    \label{fig:mederr}
\end{figure}
%%%%%%%%%%%%%%%%%%%%%%%%%%%%%%%%%%%%%%%%%%%%%%%%%%%%%%%%%%%%%%%%%%%%

We explore this in more detail in Fig.~\ref{fig:mederr}, where we show for our simulated galaxies the ratio of the estimated and the true mass as a function of the number of tracers used to calculate $M_{1/2}$ from Eq.~\ref{eq:wolf}. Every galaxy in our catalog is used at each point along the x-axis, using in each case a new random realization of $N_{GC} = 2, 3, ...N$ GCs, with N being the maximum number of of \textit{candidate} GCs that were tagged for a given galaxy. Note that this is different from the procedure in Fig~\ref{fig:massest}, where each galaxy is included only once using their \textit{realistic} number of GCs. This is done to explicitly check how the number of available tracers affects/imporves the mass estimates keeping everything else fixed in the sample. 

%\textcolor{cyan}{Every galaxy in our catalog is used at each point along the x-axis}, using in each case a new random realization of $N_{\rm GC} = 2, 3,..N$ GCs, with N being the maximum number of {\it candidate} GCs that were tagged for a given galaxy. \textcolor{cyan}{This is done to improve statistics for each $N_{\rm GC}$ bin.}Note that this is different from the procedure in Fig.~\ref{fig:massest}, where each galaxy is included only once using their {\it realistic} number of GCs.

The upper panel in Fig.~\ref{fig:mederr} corresponds to velocity dispersion estimates using r.m.s (magenta) and biweight (orange), where for each galaxy we calculate $\sigma$ as the median of the PDF corresponding to $10^5$ sub-samplings of GCs with a given number of N tracers (similar to Fig.~\ref{fig:pdfs}). The bottom panel of Fig.~\ref{fig:mederr} shows a similar exercise but using MCMC with flat (purple) and Jefferson (teal) priors. Due to computational demands, MCMC estimation corresponds, for each galaxy, to a single realization of N-tracers using $10^5$ iterations across parameter ($\sigma$, $\langle v \rangle$) parameter space as done in Fig.~\ref{fig:pdfs_mcmc}, where $\langle v \rangle$ is the assumed average 1D velocity.

We find some interesting trends. First, the accuracy of the mass estimator depends strongly on the number of tracers but not on the galaxy mass. Different shaped symbols in Fig.~\ref{fig:mederr} indicate $5$ stellar mass ranges of our galaxies, as quoted in the legend, but symbols tend to overlap suggesting little to no dependence on mass. Second, the r.m.s estimates recover the mass within 10\% for $\sim 5$-$10$ GCs while biweight requires $15$-$20$ GCs to recover the mass with the same accuracy. MCMC with a flat prior converges more slowly, needing 30-40 GCs to recover the mass within 10$\%$ while Jeffreys prior brings the requirements down to $10$-$15$ GCs for a 10\% accuracy.

Another interesting point to highlight from Fig.~\ref{fig:mederr} is the systematic deviations on the mass estimates for the different $\sigma$ measurements. Whereas $\sigma_{\rm MCMC,f}$ will tend to overestimate the mass when using fewer than $\sim 30$ GCs (see purple symbols), $\sigma_{\rm rms}$ and $\sigma_{\rm biweight}$ will underestimate the mass in the case of a low number of tracers (magenta and orange symbols). Noteworthy, using Jeffreys priors for the MCMC method can help mitigate the overestimation bias when the number of tracers is small $N_{\rm GC} \leq 5$ (green symbols), with significantly improved accuracy compared to assuming a flat prior. For larger number of tracers the assumptions on the prior do not have a significant impact. 

Our results in Fig.~\ref{fig:mederr} may be used as calibrations to improve the accuracy of mass estimation in observations of galaxies with a low number of GCs. 
We model the ratio M$_{1/2}$/M$_{\rm true}$ in for $\sigma_{\rm MCMC}$ and $\sigma_{\rm biweight}$ as: 

\begin{equation}
    \rm log \bigg( \frac{\rm M_{1/2}}{\rm M_{\rm true}} \bigg) = \frac{a}{(\rm log (\rm N_{\rm GC} ) + \rm c)^{\rm b}}
    \label{eq:errmcmc}
\end{equation}

\noindent
where $a$, $b$ and $c$ are the best fit to the medians for each method in Fig.~\ref{fig:mederr}, and the results are shown with dashed purple and orange lines for MCMC (flat prior) and biweight, respectively. Following a similar procedure, we use the following function to describe the accuracy of mass estimation when using r.m.s velocity dispersions: 

\begin{equation}
    \frac{\rm M_{1/2}}{\rm M_{\rm true}} = \frac{\rm a}{\rm N_{\rm GC}^{\rm b}} + c
    \label{eq:err_rms}
\end{equation}

\noindent
Our best fit values $a$, $b$ and $c$ for the three velocity dispersion estimates are summarized in Table~\ref{tab:err_params}. We hasten to add that the corrections for the MCMC case will  depend on the shape of the prior. For example, in the case of the Jeffreys prior the correction to the median is roughly well described by a constant upwards shift factor of $\sim 1.5$, albeit with a significant object to object scatter.

%%%%%%%%%%%%%%%%%%%%%%%%%%%%%%%%%%%%%%%%%%%%%%%%%%%%%%%%%%%%%%%%%%%%
\begin{table}
    \centering
    \begin{tabular}{|c|c|c|c|c|}
    \hline 
    Estimate & a & b & c & Equation \\ 
    \hline 
    $\sigma_{\rm rms}$ & -1.535 & 1.057 & 9.963 & \ref{eq:err_rms} \\ 
    \hline 
    $\sigma_{\rm biweight}$ & -1.956 & 1.110 & 1.004 & \ref{eq:err_rms} \\ 
    \hline 
    $\sigma_{\rm MCMC,f}$ & 0.097 & 1.908 & -0.024 & \ref{eq:errmcmc} \\ 
    \hline 
    \end{tabular}
    \caption{Values for the parameters in equations \ref{eq:errmcmc} and \ref{eq:err_rms} for each of the $\sigma_{\rm los}$ estimates.}
    \label{tab:err_params}
\end{table}
%%%%%%%%%%%%%%%%%%%%%%%%%%%%%%%%%%%%%%%%%%%%%%%%%%%%%%%%%%%%%%%%%%%%

\subsection{Impact of tidal stripping}
\label{ssec:tidal}

As an important application of our GC catalog, we can use the cosmological simulations of galaxies within realistic cluster environments to quantify how much tidal stripping might affect the accuracy of mass recovery techniques similar to Eq.~\ref{eq:wolf} using GCs as tracers. Since Jeans modeling assumes the system to be in equilibrium, tidal stripping could potentially bias the results or cause the mass estimators to perform less accurately for significantly stripped and disturbed systems, as suggested by \citet{Smith2013} in the context of galaxy harassment.

\begin{figure}
    \centering
    \includegraphics[width=\columnwidth]{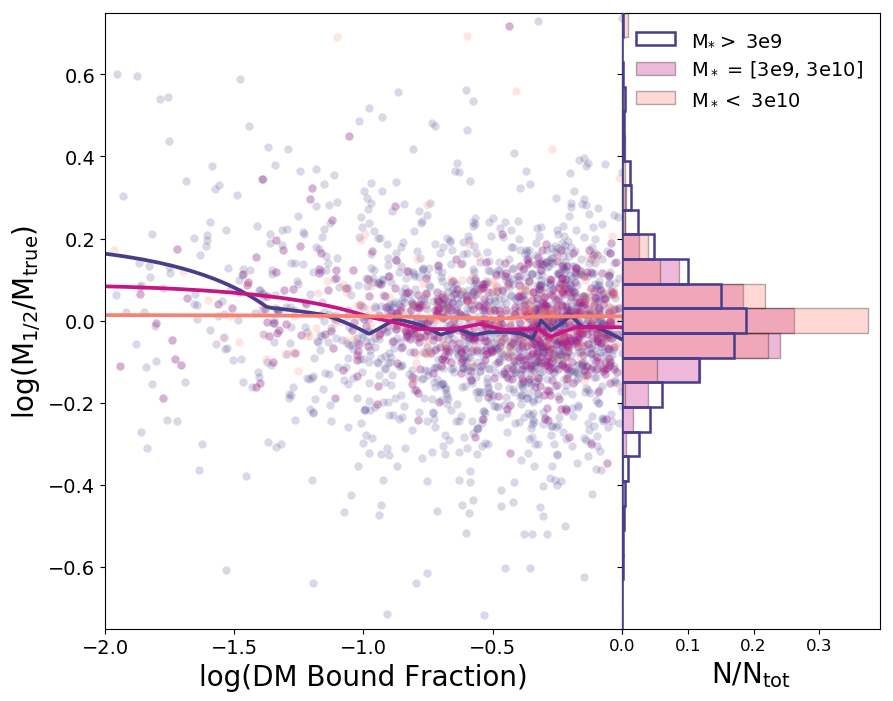}
	\caption{Ratio between the estimated mass using GCs to the true mass in simulated galaxies as a function of the amount of tidal stripping experienced. We show results for $\sigma_{\rm MCMC}$ but similar results applies to the other definitions. The fraction of bound DM mass (x-axis) is calculated as the $z=0$ dark matter mass compared to the infall value. In general, the median of $M_{1/2}/M_{\rm true}$ of the sample (solid lines) shows little dependence on the remaining dark matter bound mass fraction, providing confidence on mass estimation methods even within the tidal environment of clusters. We find no significant trend with the galaxies stellar mass (see different colors).} 
    \label{fig:tidalstrip}
\end{figure}

We find that, contrary to these expectations, Eq.~\ref{eq:wolf} performs {\it on average} extremely well even in cases with significant mass loss. Fig.~\ref{fig:tidalstrip} shows the ratio of recovered mass using GCs, $M_{1/2}$, to the real mass from the simulation, $M_{\rm true}$, compared to the fraction of dark matter mass that is still bound (DM bound fraction), which we define to be the ratio of the present day dark matter mass of a galaxy to that at its time of infall. Different colors correspond to different stellar mass ranges for our galaxies and we find no significant trend with mass. This test uses $\sigma_{\rm MCMC}$ (with flat priors) to estimate the velocity dispersion of each galaxy using their realistic number of GCs in our catalog, but we have explicitly checked that the conclusions do not change if we use either $\sigma_{\rm MCMC,j}$, $\sigma_{\rm rms}$ or $\sigma_{\rm biweight}$.

%%%%%%%%%%%%%%%%%%%%%%%%%%%%%%%%%%%%%%%%%%%%%%%%%%%%%%%%%%%%%%%%%%%%
%%%%%%%%%%%%%%%%%%%%%%%%%%%%%%%%%%%%%%%%%%%%%%%%%%%%%%%%%%%%%%%%%%%%

\begin{figure}
    \centering
    \includegraphics[width=\columnwidth]{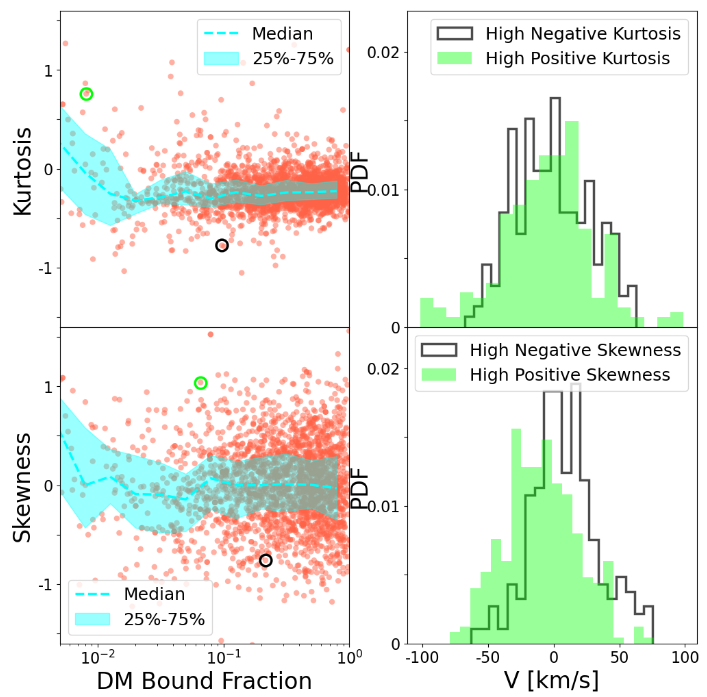}
	\caption{\textit{Left}: Kurtosis (upper panel) and skewness (lower panel) as a function of the fraction of dark matter mass retained at z = 0 compared to that at infall. The median and 25-75 percentile range are shown by the cyan dotted lines and shading respectively. While the scatter of the skewness with respect to the DM bound fraction remains relatively constant, the scatter of the kurtosis increases as the DM bound fraction decreases. We also see an increase in the median of both the kurtosis and skewness with a decrease in DM bound fraction, suggesting that tidal stripping might induce a bias towards higher values. \textit{Right}: Examples of non-Gaussian velocity distributions for extreme values of Kurtosis (upper panel) and extreme values of skewness (lower panel). The color of the histograms corresponds to the same colored circled points on the right panels.}
    \label{fig:kurt_skew}
\end{figure}

A more detailed look at tidally stripped systems might reveal, however, important trends affecting the shape of the velocity distribution of tagged GC candidates. Fig.~\ref{fig:kurt_skew} quantifies the kurtosis (top) and skewness (bottom) of the line-of-sight velocity distribution of GCs for each of our galaxies as a function of their retained dark matter mass fraction. A perfectly Gaussian function corresponds to kurtosis and skewness being both consistent with zero. The cyan line and shaded regions correspond to the median and $1$ $\sigma$ scatter of the sample at fixed bound mass fraction. 

%%%%%%%%%%%%%%%%%%%%%%%%%%%%%%%%%%%%%%%%%%%%%%%%%%%%%%%%%%%%%%%%%%%%
%%%%%%%%%%%%%%%%%%%%%%%%%%%%%%%%%%%%%%%%%%%%%%%%%%%%%%%%%%%%%%%%%%%%

Although GCs might {\it reasonably} be well described by Gaussians, our sample of candidate GCs systems show a systematic trend to negative kurtosis (median $\sim -0.3$ for objects with no significant stripping) and overall significant scatter in both kurtosis and skewness. Histograms on the right panels of Fig.~\ref{fig:kurt_skew} show examples of the shape of the velocity distribution of GCs for galaxies with either high or low skewness or kurtosis. 

These deviations from Gaussianity might be more common for galaxies under severe tidal stripping (dark matter bound fraction lower than a few percent), which exhibit a bias towards higher values of kurtosis and skewness and increased scatter, especially in kurtosis. These results are important in light of the common-practice assumption of Gaussianity to estimate the uncertainties in the velocity dispersion of GCs in observational studies. How can skewness and kurtosis affect the calculated confidence intervals?

%%%%%%%%%%%%%%%%%%%%%%%%%%%%%%%%%%%%%%%%%%%%%%%%%%%%%%%%%%%%%%%%%%%%
\begin{figure}
    \centering
    \includegraphics[width=\columnwidth]{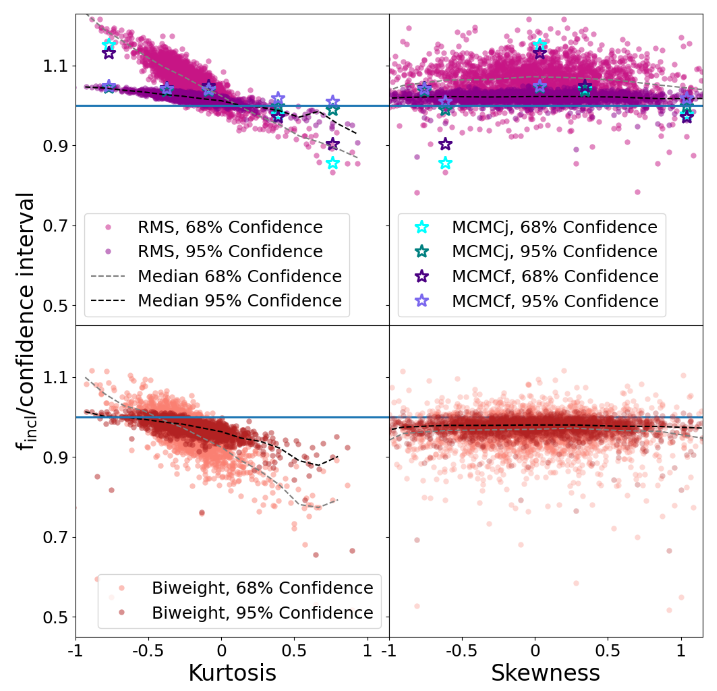}
	\caption{Correction to the Gaussian confidence intervals as a function of kurtosis (left) and skewness (right) of the distribution of candidate GCs associated to our galaxy sample. Estimates are based on $1000$ resampling of $10$ GCs (see text for details). The top and bottom row correspond to r.m.s and biweight estimates, respectively. Overlaid on the top row, we show with starred symbols the same calculation but using both MCMC methods for the four systems highlighted in Fig.~\ref{fig:tidalstrip} plus our fiducial dwarf in the top panel of Fig.~\ref{fig:introfig}. Different colors correspond to $68\%$ and $95\%$ confidence intervals, as labeled. Thin dashed lines highlight the median correction at fixed kurtosis or skewness. Non-Gaussianities may have a significant (and systematic) impact on accuracy estimates, in particular for high/low kurtosis values.}
    \label{fig:error_correc}
\end{figure}
%%%%%%%%%%%%%%%%%%%%%%%%%%%%%%%%%%%%%%%%%%%%%%%%%%%%%%%%%%%%%%%%%%%%

Confidence intervals are formally defined as the probability that the true variance of a given sample (in this case, all GCs candidates) lays within the confidence interval of the variance of a random subsample (for example, the realistic GC number) drawn from such parent distribution. If the underlying population is non-Gaussian, that probability would be expected to change and therefore confidence intervals can be over- or under-estimated. We show this in Fig.~\ref{fig:error_correc} using the r.m.s method (circles). For each galaxy we generate $1000$ resampling of $10$ GCs and compare the re-calculated confidence intervals to that of a Gaussian distribution. See Appendix~\ref{app:gaussian} for more details. 

We find that variations in kurtosis result in well-defined trends for the non-Gaussian confidence intervals (top left panel Fig.~\ref{fig:error_correc}). In GC systems with intrinsic negative kurtosis, the confidence intervals are overestimated, meaning that the probability of finding the true variance within the computed confidence interval is actually larger than the case of a Gaussian distribution. For such systems, the observed value is actually {\it more} accurate than expected in a Gaussian case. The opposite is true for systems with positive kurtosis, where confidence intervals are underestimated. The scale of the effect varies with the confidence level being considered, varying from $10$-$20\%$ for the 68\% percentile (magenta) to $\sim 5\%$ for 95\% confidence level (salmon). 

Given the overall bias of our GC population in Fig.~\ref{fig:kurt_skew} towards negative kurtosis, current uncertainties calculated in observations might actually be on the conservative side and constrains actually tighter than currently estimated. This, however, changes for systems under severe tidal disruption, expected to show more often positive kurtosis values that could result on confidence intervals being currently underestimated in the literature. 

A similar exercise sorting our galaxies by their skewness (top right panel of Fig.~\ref{fig:error_correc}) shows no significant dependence of the correction to confidence intervals with this parameter. Note that although these results were derived for r.m.s estimates, examples calculated from MCMC are consistent with these results (starred symbols). For completeness, we also show the correction levels for biweight velocity dispersion (see bottom panels of Fig.~\ref{fig:error_correc}) which agree well with those calculated for r.m.s.

We conclude that although the overall velocity dispersion and dynamical mass estimates perform remarkably well {\it on average}, even under severe tidal disruption, in individual objects, kurtosis might be an important factor to consider when reporting confidence intervals in observations. This seems roughly independent of the particular method used to calculate the velocity dispersion, at least among the three explored here: r.m.s, biweight and MCMC. Unfortunately, estimating kurtosis or skewness in a sample with only a handful of GCs is challenging. Our theoretical results should be interpreted mostly as a warning that large deviations from Gaussianity may occur and would have a sizable impact on the estimated confidence intervals. This may have important consequences when dealing with systems where tidal disruption may be suspected to be important, as is the case of some ultradiffuse dwarf galaxies.
    
%%%%%%%%%%%%%%%%%%%%%%%%%%%%%%%%%%%%%%%%%%%%%%%%%%%%%%%%%%%%%%%%%%%%
\begin{figure*}
    \centering
    \includegraphics[scale = 0.8]{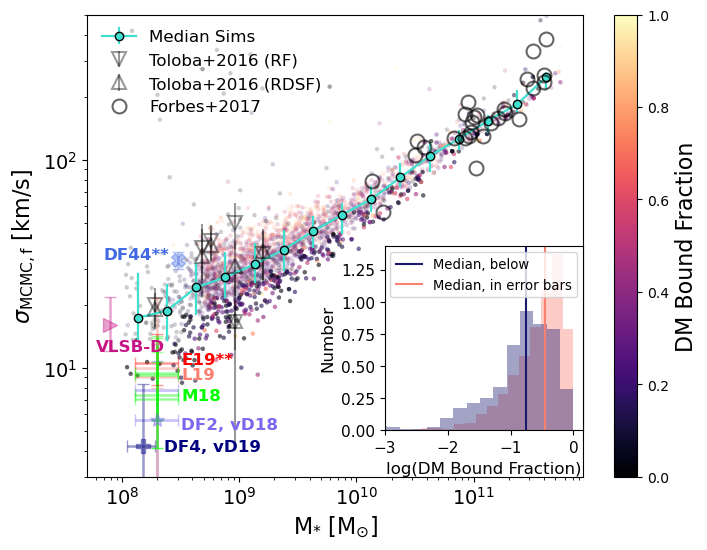}
	\caption{Predicted line-of-sight velocity dispersion of simulated GCs, $\sigma_{\rm MCMC,f}$, as a function of host galaxy stellar mass in our simulations, color-coded by the DM bound fraction, M$_{\rm DM, \rm z=0}$ / M$_{\rm DM, \rm inf}$ of each galaxy. The median trend (cyan) is in good  agreement with observational constraints from SLUGGS on the high mass end (open circles) and also dE galaxies in Virgo \citet{Toloba2016} using both their RF (rotation fit) and RDSF (rotation and dispersion simultaneous fit) methods. The color gradient in the simulated points shows that at fixed $M_*$ galaxies that have experienced more tidal disruption have the lowest GC velocity dispersion (darker symbols). We highlight this by selecting all simulated galaxies below and within $1\sigma$ scatter of the mean relation (see points with higher opacity and those within the cyan error bars) and plotting their distribution of retained dark matter mass in the small inset. These low $\sigma_{\rm MCMC,f}$ galaxies have retained only $17\%$ (median) of their initial dark matter mass compared to about $36\%$ of that for galaxies within 1 $\sigma$ of the median. Interestingly, ultradiffuse galaxies with similar stellar masses show a wide range of velocity dispersion. Data for only 4 UDGs are available in this mass range, VLSB-D \citep{Toloba2018}, DF44** \citep{vanDokkum2019c}, DF4 \citep{vanDokkum2019a} and several estimates for DF2: vD19 \citep[lavander, ][]{vanDokkum2019c}, M18 \citep[green, ][]{Martin2018}, L19 \citep[orange, ][]{Laporte2019}, E19** \citep[red, ][]{Emsellem2019}. (Double asterisks indicate velocity dispersion of the stellar component and not from GCs.) Examples like DF2 and DF4 sit at the lowest bounds of velocity dispersion with $\sigma \leq 10$ km/s. These results hint at tidal disruption as a possible formation path for objects like DF2 and DF4. In particular, some of our simulated galaxies overlap with the constraints for DF2.} 
    \label{fig:obs}
\end{figure*}
%%%%%%%%%%%%%%%%%%%%%%%%%%%%%%%%%%%%%%%%%%%%%%%%%%%%%%%%%%%%%%%%%%%%

\section{Dark matter content in dwarfs estimated from the kinematics of GCs}
\label{sec:df2}

Dwarf galaxies place constraints and challenges to the cosmological $\Lambda$CDM scenario and, with it, an opportunity to test theoretical predictions and validate (or falsify) the cosmological model. One of the basic predictions of galaxy formation models in the $\Lambda$CDM framework is that dwarf galaxies inhabit relatively massive halos. A number of observational efforts have therefore focused on measuring the dark matter content in dwarfs. In the case of cluster dwarfs, which are in their majority gas poor and of low surface brightness, GCs are often the best dynamical tracers given their luminosity and extended spatial distribution. 

Studies of the kinematics of GCs in several dE galaxies in Virgo have revealed a wide range of velocity dispersion for GCs in  $M_* \sim 10^9$\msun\ dwarfs \citep{Toloba2016}. However, other studies targeting ultra-diffuse dwarfs have revealed a much wider GC velocity dispersion range, including the detection of some UDGs where $\sigma_{\rm GC}$ is so low that, at face value, it suggests systems that are ``dark matter free" \citep{vanDokkum2018a, vanDokkum2018b, vanDokkum2019a, Toloba2018}. This result offers vital clues to our understanding of the formation paths of UDGs in clusters. 

We use our tagged GC catalog in Illustris to study the population and kinematics of GCs predicted for dwarfs in clusters like Virgo. Fig.~\ref{fig:obs} shows the $\sigma_{\rm MCMC,f}$ of GCs as a function of the stellar mass in our simulated cluster galaxies. We choose $\sigma_{\rm MCMC,f}$ to facilitate the comparison with observational data. The median of the simulated relation is indicated in cyan, with vertical error bars corresponding to the r.m.s scatter.

In the dwarf regime (i.e., $M_*<10^9\, M_\odot$) our estimates of $\sigma_{\rm GC}$ agree well with those of dEs in Virgo \citep[data from ][ shown in grey triangles in Fig.~\ref{fig:obs}]{Toloba2016}. This is encouraging, since the GC tagging method relies on observations and calibrations done at higher masses, and the power-law relation between halo mass and GC mass is an extrapolation over this mass range. Moreover, the tagging is done at the moment of infall into the cluster and not at present day, making this comparison mostly a prediction of the model. Furthermore, it is reassuring that the velocity dispersion predicted for more massive ellipticals agree well with constraints from the SLUGGs survey \citep[see open gray circles, ][]{Forbes2017c}.

Our calculations have so far not included the effect of individual errors in the measured velocity of each GC.
In the dwarf galaxy regime, observations typically have individual errors of order $3$-$10$ km s$^{-1}$ per GC \citep{Toloba2016, Toloba2018, vanDokkum2018a}. We have checked that adding random Gaussian errors with $10$ km s$^{-1}$ to our GC velocities only increases $\sigma_{\rm MCMC,f}$ on average by $\sim 20\%$ on our lowest velocity dispersion objects, with increasingly smaller effect towards more massive systems. For instance, in galaxies with $\sigma_{\rm MCMC,f} \sim 25$ km s$^{-1}$, the MCMC velocity dispersion calculated assuming $10$ km s$^{-1}$ errors exceeds that without errors by $\sim 5\%$ (median, see Fig.~\ref{fig:mcmc_err}). The overestimation is even smaller if we assume random errors with amplitude $5$ km s$^{-1}$ instead (see Appendix~\ref{app:errors} for more details). 

Simulated galaxies in Fig.~\ref{fig:obs} are colored by their retained (bound) dark matter fraction, calculated, defined as before as the ratio present-day dark matter mass given by {\sc subfind} compared to that at the moment of infall. We find a clear gradient of $\sigma_{\rm MCMC}$ at fixed $M_*$, where galaxies with high GC velocity dispersion tend to retain most of their dark matter mass while low $\sigma_{\rm MCMC}$ values are dominated by galaxies that have lost more than $80\%$ of their dark matter mass. To highlight this we show in the inset panel the distribution of bound dark matter mass fraction for all galaxies that deviate by more than one-sigma below the median relation (included points are highlighted with a higher symbol opacity). Galaxies this far down in velocity dispersion have retained typically only $17\%$ of their initial dark matter halo. 

Can tidal stripping explain the low GC velocity dispersion found in some UDGs like DF2? We show in Fig.~\ref{fig:obs} several measurements for the velocity dispersion of this dwarf as determined by different teams using slightly different assumptions \citep{Martin2018, Laporte2019, vanDokkum2018a}. Interestingly, we find a few simulated dwarfs with $\sigma_{\rm MCMC}$ consistent with the upper end of the range measured in the literature for DF2. These objects in our simulations seem significantly tidally stripped (dark color points), in agreement with the arguments discussed in the previous paragraph. 

These results are intriguing, since tidal disruption has been proposed as one of the mechanisms that may transform normal galaxies into UDGs in clusters \citep{Carleton2019, Sales2020, Leigh2020, Maccio2020}, and some observational evidence for the case of stripping has recently been presented \citep{Montes2020}. Although the simulations do not have the resolution to follow the morphological changes of these galaxies, our results suggest that the same tidal transformation might lead to velocity dispersions as low ($\sigma_{\rm GC} \sim 10$ km/s)  as that observed for GCs around DF2.  

We note that the stellar mass for DF2 is estimated to be about $M_* \sim 2 \times 10^8$\msun\; \citep{vanDokkum2018a}, which is below our cutoff $M_* \geq 5 \times 10^8$\msun\; to tag GCs onto infalling halos. This means that our sample at these small masses includes only dwarfs that were more massive in the past (and therefore fulfilled our cut of $5 \times 10^8$\msun\; for the initial tagging). From this perspective, it is not surprising the tidal origin of our identified DF2-analogs. However, it is interesting to find objects with GC velocity dispersions as low as DF2 in our simulated clusters. 

\begin{figure}
    \centering
    \includegraphics[width = \columnwidth]{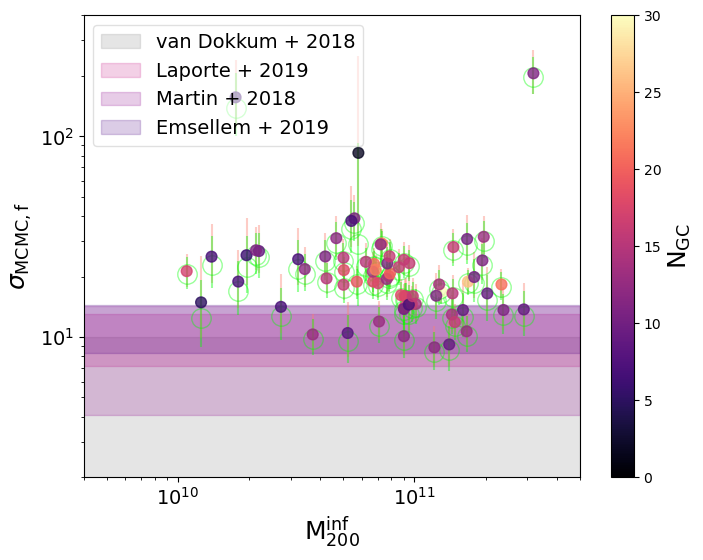}
	\caption{GCs velocity dispersion $\sigma_{\rm MCMC,f}$ for simulated dwarfs in the stellar mass range comparable to DF2, $M_*=1$-$3 \times 10^8$\msun\; as a function of their infall virial mass $M_{200}^{\rm inf}$. The vertical error bars correspond to the 68 percent confidence interval for the MCMC velocity dispersion estimate (flat priors are shown in full symbols, open green circles show Jeffreys priors). Several of our simulated objects show GCs velocity dispersion that are compatible with the upper end of observational estimates for DF2 (see shaded areas). These kinematic analogs of DF2 have 3-17 GCs still bound at $z=0$ (color bar) in good agreement with the $\sim 10$ GCs currently known for DF2. Our simulations suggest that DF2-like objects may have infell as dwarf halos with $M_{200}^{\rm inf}=0.3$-$3 \times 10^{11}$\msun\; loosing more than $90$\% of their dark matter mass at present day.}
    \label{fig:df2}
\end{figure}

To better assess the dark matter halos inhabited by DF2 candidates, we select all our simulated dwarfs in the stellar mass range $M_*=[1$-$3]\times 10^8$\msun\; and show their present-day $\sigma_{\rm MCMC, f}$ of GCs as a function of their infall virial mass in Fig.~\ref{fig:df2}. Here each simulated dwarf is color-coded by the number of GCs retained. We find that these ``DF2-analogs'' have between 3-30 GCs, in good agreement with the 9-11 observed GCs around DF2. We also show that the assumption of flat (full symbols) or Jeffreys (green open circles) priors do not qualitatively change our results \citep[in agreement with the conclusions of][]{Martin2018}. For comparison, the shaded horizontal regions in Fig.~\ref{fig:df2} indicate the observational estimates of the velocity dispersion of DF2 GCs according to various authors.

%%%%%%%%%%%%%%%%%%%%%%%%%%%%%%%%%%%%%%%%%%%%%%%%%%%%%%%%%%%%%%%%%%%%
%%%%%%%%%%%%%%%%%%%%%%%%%%%%%%%%%%%%%%%%%%%%%%%%%%%%%%%%%%%%%%%%%%%%

Although the majority of our simulated dwarfs have higher $\sigma_{\rm MCMC,f}$, there are a handful of objects that overlap with the uncertainty range from \citet{Martin2018} and \citet{Laporte2019}. These objects had infall virial masses consistent with dwarf halos in the range $M_{\rm 200}^{\rm inf} \sim [0.3$-$3] \times 10^{11}$\msun, comparable to that estimated for, for example, the Large Magellanic Cloud (LMC) in the Milky Way. For these objects, the number of GCs predicted by our tagging method (3-17) agrees well with the $\sim 10$ GCs found associated to DF2. 

We note that there are still significant uncertainties in the measurement of the GC velocity dispersion in DF2. If lower values are proven more accurate, this would place DF2 close to the even more ``dark-matter-free'' UDG DF4 (as least as suspected from its stellar velocity dispersion).  Additional formation mechanisms might be needed to explain the very low dark matter density in these extreme class of objects, such as a ``tidal-dwarf'' origin \citep[see; e.g.,][]{Zwicky1956,Schweizer1978,Mirabel1992}. The presence of dark matter cores driven by dark matter self-interactions or by baryonic feedback could represent a possible solution to this problem. Furthermore, higher resolution simulations would be needed to resolve such low dark matter contents. 

Considering non-Gaussianities and the corrections to confidence intervals explored in Sec.~\ref{ssec:tidal}, if the GC population  of DF2 had positive kurtosis, then the uncertainty ranges estimated should be revised upwards, which would help alleviate the tension. In particular, the estimate from  \citet{vanDokkum2018a} is based in $\sigma_{\rm biweight}$, for which we find a systematic underestimation of the confidence intervals compared to a Gaussian in the case of positive kurtosis (see Fig.~\ref{fig:error_correc}). Assuming a correction level of $15\%$ (corresponding to an intrinsic kurtosis $\sim 0.75$) DF2 could increase the upper limit of the $90\%$ confidence interval from $10.5$ km/s to $12.1$ km/s, bringing it closer to other estimates\footnote{We note that the procedure in \citet{vanDokkum2018a} (and reproduced by \citet{Martin2018}) is slightly different than derived in Sec.~\ref{ssec:tidal}, since they jointly estimate confidence intervals and intrinsic velocity dispersion in a single step.}.

We conclude that it is indeed possible that DF2 may have formed as the result of a normal dwarf halo that has been stripped of more than $90\%$ of its mass. Low surface brightness stellar tails, elongated morphology or evidence of rotation for its GC system (such as that found by \citet{Lewis2020}) could help confirm its tidal nature but their absence will not conclusively rule out this formation path. This highlights the urgent need for more observational campaigns targeting the kinematics of GCs around UDGs to more robustly constrain their global dark matter content. 

\section{Summary}
\label{sec:concl}

We use a catalog of GCs tagged onto the cosmological hydrodynamical simulation Illustris to study the accuracy of dynamical mass estimates based on the radial extension and line-of-sight velocities of GCs systems. In particular, we analyze the GC system of satellite galaxies in $9$ simulated galaxy clusters with virial mass $M_{200} \sim 10^{14}$\msun. Our sample consist of 3777 galaxies in the mass range $M_*=10^8$-$6 \times 10^{11}$\msun.

We find that mass estimators of the form M$ \propto \sigma^2 \rm r$ do a remarkably good job at estimating mass when using GCs as tracers, specially when having $10$ or more GCs. For galaxies that have a smaller number of GCs with measured kinematics, the particular definition of velocity dispersion used may systematically bias the results. Using the r.m.s and biweight velocity dispersion \citep{Girardi2008, Veljanoski2014, vanDokkum2018a} tends to underestimate the dynamical mass, while other methods used in the literature such as $\sigma_{\rm MCMC}$  \citep{Widrow2008, Hogg2010, Toloba2016, Martin2018} tend, on the contrary, to overestimate masses for low $N_{\rm GC}$. In the case of MCMC the shape of the prior may play an important role, with Jeffreys prior resulting on a lower bias compared to a flat prior. We provide fitting formulas in Eq.~\ref{eq:err_rms} and ~\ref{eq:errmcmc} that might help  correct for these effects in observational samples with less than $10$ GC tracers. 

Surprisingly, the accuracy of the recovered mass estimation depends little on the level of tidal disruption suffered by the galaxy, indicating that satellite galaxies in clusters are, in their majority, in state of quasi-equilibrium as soon as they move away from their pericenters \citep[see; e.g.,][]{Penarrubia2009}. Our results provide strong support for the use of GC kinematics to estimate dynamical masses even in high density environments such as clusters. A word of caution is necessary in the case of systems with significant tidal stripping, where significant deviations in kurtosis may arise as a consequence of the tidal disruption effects and may impact the estimated confidence intervals. 

We use our results to compare the dark matter content of cluster dwarfs predicted in our simulations with available observational constraints in nearby clusters and groups. We find good agreement with the median and scatter measured for dwarf ellipticals in Virgo. Moreover, we find that tidal disruption creates scatter in the measured $M_*$-$\sigma_{\rm GC}$ such that at a fixed $M_*$, smaller velocity dispersions correlate with larger mass losses to tidal disruption. On average, galaxies that lay below the median relation by $1\sigma$ have lost $\sim 83\%$ of their infall dark matter mass. 

In observations, there is  large scatter in the velocity dispersion of GCs for dwarfs with $M_* \leq 10^9$\msun, with the most extreme outliers being ultra-diffuse galaxies DF2 and DF4. These UDGs have estimated GC velocity dispersions in the range $7-10$ km/s (DF2) and $\sigma \sim 4.2$ km/s (DF4) suggesting that they are extremely dark matter deficient. Interestingly, we identify a set of dark matter poor DF2-analogs in our simulation that have similar stellar masses, 3-17 bound GCs, and a velocity dispersion of those GCs $\sigma_{\rm MCMC} \sim 10$ km/s, consistent with the upper envelope of measured values for DF2.  The progenitors of these DF2 analogs fell into the cluster as dwarf halos with $M_{200}^{\rm inf} =[0.3$-$3] \times 10^{11}$\msun but have lost more than 90\% of their mass to tidal disruption. Interestingly, tidal disruption has also been proposed as possible mechanism to form UDGs in clusters \citep{Carleton2019, Sales2020, Leigh2020, Maccio2020, Montes2020}. Our results suggest that the same mechanism may be able to explain simultaneously the ultra-diffuse nature {\it and} low GC velocity dispersion in objects alike DF2 within $\Lambda$CDM.

Although we do not find systems with velocity dispersions as low as that inferred for DF4, we are limited by numerical resolution in our Illustris sample. The small number of UDGs with available kinematical data does not allow for a proper evaluation of how common or rare dark-matter poor dwarfs like DF2 and DF4 might be, or their dependence on the environment or host mass. While systematic photometric studies of UDGs and their GCs in nearby groups and clusters are starting to become available \citep[e.g., in the Virgo cluster][]{Lim2020}, adding spectroscopic data to constrain their stellar and GC kinematics would represent the most promising avenue towards a better understanding of how UDGs form.

Explaining the large scatter in the dark matter content of dwarf galaxies is one of the outstanding challenges in the $\Lambda$CDM model. While the rotation curves of gas-rich dwarfs have revealed a wide variety of dark matter distribution in field dwarfs, GCs are starting to reveal a similarly rich complexity for gas-poor dwarfs in groups and clusters. As we look forward to larger datasets with available GC kinematical constraints for early-type dwarfs, our results validate the use of GCs as efficient dynamical mass estimators even in the case of a modest number of GCs with measured kinematics.

\section*{Acknowledgements}
We would like to thank useful and stimulating discussions with Matt Walker and Aaron Romanowsky over this draft. JED and LVS acknowledge support from NASA-ATP-80NSSC20K0566 and NSF-CAREER-1945310 grants. JED is grateful to the DAAD short-term research grant and the Max Planck Institute for Astrophysics in Garching, Germany. 

\section*{Data Availability}
This paper is based on merger trees and halo catalogs from the Illustris Project \citep{Vogelsberger2014b, vogelsberger2014a}. This data is available publically at \url{https://www.illustrisproject.org/}. The catalog of GCs and other products of this analysis may be shared upon request to the corresponding author if no further conflict exists with ongoing projects.

%%%%%%%%%%%%%%%%%%%%%%%%%%%%%%%%%%%%%%%%%%%%%%%%%%

%%%%%%%%%%%%%%%%%%%% REFERENCES %%%%%%%%%%%%%%%%%%

% The best way to enter references is to use BibTeX:

\bibliographystyle{mnras}
\bibliography{gcs} % if your bibtex file is called example.bib

% Alternatively you could enter them by hand, like this:
% This method is tedious and prone to error if you have lots of references
%\begin{thebibliography}{99}
%\bibliography{gcs}
%\end{thebibliography}

%%%%%%%%%%%%%%%%%%%%%%%%%%%%%%%%%%%%%%%%%%%%%%%%%%

%%%%%%%%%%%%%%%%% APPENDICES %%%%%%%%%%%%%%%%%%%%%

\appendix

%% ----------------------------------------------------------------
\section{Velocity Dispersion Measurements}
\label{app:sigma}

Here is a more detailed discussion of the calculations of each of the methods used to calculate velocity dispersion in this work.

\begin{itemize}
\item {\bf The r.m.s dispersion, $\sigma_{\rm rms}$}
    
This methods assumes that the underlying velocity distribution is Gaussian, and it is calculated using:
\begin{equation}
    \sigma_{\rm rms} = \sqrt{\frac{\sum_{\rm i}^{\rm N}{(\rm v_{\rm i} - \bar{v})^2}}{N}}
    \label{eqn:srms}
\end{equation}
\noindent where $N$ is the number of GC tracers, v$_{\rm i}$ are the individual velocities of the GCs, and $\bar{\rm v}$ is the center of mass velocity of the galaxy the GCs are associated with.

We first use this calculation to perform 3-$\sigma$ clipping of the GC candidate particles from which we later draw our realistic sample of GCs. This removes most of the GC particles within the cutoff radius that belong to the intracluster population and thus would contaminate our sample.\\

\item {\bf The biweight velocity dispersion $\sigma_{\rm biweight}$}

This method does not assume an underlying Gaussian velocity distribution and instead assigns different set of weights to each velocity measurement, where larger weight values are given to velocities closer to the median of the distribution. This method is advantageous for  highly contaminated samples of tracers, where the biweight estimation downweights possible outliers or contaminants making them less influential in the final $\sigma$ estimate compared to the simpler r.m.s calculation. 

As introduced in \citep{Beers1990}, to calculate the biweight estimation of scale, we first need to calculate the mean absolute deviation (MAD):
\begin{equation}
    \rm MAD = \rm median(|\rm v_{\rm i} - \rm M|)
    \label{eq:mad}
\end{equation}

\noindent 
where v$_{\rm i}$ are the individual velocities and M is the median of those velocities. Next, we calculate u$_{i}$, the weight associated with each velocity following:

\begin{equation}
    \rm u_{\rm i} = \frac{\rm v_{\rm i} - M}{\rm cMAD}
    \label{eq:ui}
\end{equation}

\noindent 
where c is the ``tuning'' parameter, which is to be set to 9 according to \citet{Beers1990}. The biweight estimation of scale is then given by:

\begin{equation}
    \sigma_{\rm biweight} = \frac{\rm N^{1/2} [ \sum_{|\rm u_{\rm i}| < 1} {(\rm v_{\rm i} - M) (1 - \rm u_{\rm i})^4}]^{1/2}} {|\sum_{|\rm u_{\rm i}| < 1}{(1 - \rm u_{\rm i}^2)(1 - 5\rm u_{\rm i}^2)}|}
    \label{eqn:sbi}
\end{equation}

\noindent where $N$ again is the number of tracers. A minimum of $5$ tracers is required for this method to work (see \citet{Beers1990} for a brief discussion).\\

\item {\bf MCMC velocity dispersion, $\sigma_{\rm MCMC}$}

This method takes the line of sight velocity distribution and stochastically finds the best $\sigma$ and $\bar{\rm v}$ to fit a Gaussian to the distribution. The likelihood used in this estimation is given by
\begin{equation}
    \mathcal{L} = \prod_{\rm i}^{\rm N_{\rm GC}} \frac{1}{\sigma \sqrt{2\pi}} \rm exp\bigg(-0.5 \bigg(\frac{v_{\rm i} - <v>}{\sigma}\bigg)^2\bigg)
    \label{eq:likelihood}
\end{equation}

\noindent 
where v$_{\rm i}$ are the line of sight velocities of the tracers and $\bar{\rm v}$ and $\sigma$ are allowed to vary. MCMC methods tend to be quite computationally expensive and the results can depend on the shape of the prior assumed (see \citet{Martin2018} for a specific example using DF2). Here we explore two different assumptions for the priors: a uniform distribution (flat prior) and Jeffreys prior, which in the case of a Gaussian function like assumed here, corresponds to a prior distribution $\propto 1/\sigma$. 

In practice, Jeffreys prior amounts to multiplying equation \ref{eq:likelihood} by (1/$\sigma$) and has the net effect of shortening the long tails in the posterior PDF for the velocity dispersion in figures alike \ref{fig:mcmc} (see right panel in Fig.~\ref{fig:pdfs_mcmc}). The Jeffreys prior is, however, improper, which means the distribution of posterior probabilities might not necessarily integrate to 1 unless a lower limit in $\sigma$ is specified. We have used $\sigma=0.5$ km s$^{-1}$ in our calculations, but we have explicitly checked that changing that to $0.5$ or $1$ km s$^{-1}$ does not qualitatively change our results. We have confirmed that the use of the Jeffreys prior is particularly powerful for systems with small N$_{\rm GC}$, where the differences with a flat prior are most significant (see Fig.~\ref{fig:mederr}).

%%%%%%%%%%%%%%%%%%%%%%%%%%%%%%%%%%%%%%%%%%%%%%%%%%%%%%%%%%%%%%%
\begin{figure}
    \centering
    \includegraphics[scale = 0.62]{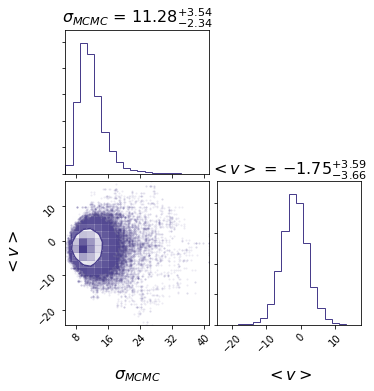}
    \caption{An illustration of the MCMC method for a randomly selected galaxy with GCs in Illustris. This pdf estimates both the velocity dispersion $\sigma$ and the expectation value $< v >$ using stochastic sampling of parameter space using 10 globular clusters. This method assumes an intrinsic error on the order of 5km/s in the velocity measurements.}
    \label{fig:mcmc}
\end{figure}
%%%%%%%%%%%%%%%%%%%%%%%%%%%%%%%%%%%%%%%%%%%%%%%%%%%%%%%%%%%%%%%

We employ the Metropolis-Hastings technique to find the posterior PDF of $\sigma_{\rm los}$, the result of which is illustrated in Figure \ref{fig:mcmc} In short, this technique involves the following:
\begin{enumerate}
    \item set initial estimates for the parameters in question. 
    \item randomly select one of those variables. Calculate the likelihood.
    \item select a random point from a Gaussian jumping distribution centered on the current value of the parameter with a dispersion set in the case of this study to 5 $\rm km$ $\rm s^{-1}$. This becomes the new value of the selected parameter. 
    \item calculate the likelihood with this new parameter value. Then, 
    \begin{itemize}
        \item if the likelihood at the new value is greater than the likelihood of the old value, we keep the new value of the parameter.
        \item if the likelihood at the new value if less than the likelihood at the new value, the if the ratio of the new likelihood to the old likelihood is greater than some random number between 0 and 1, we keep the new value of the parameter. Else, we keep the old value.
    \end{itemize}
    \item repeat until the parameter space of all variables has been sufficiently explored.
\end{enumerate}

\noindent We have illustrated this process in Figure \ref{fig:mcmc}. The corner panel shows the 2D pdf of the line of sight velocity dispersion $\sigma_{\rm MCMC}$ and the expectation value of the line of sight velocity distribution $<\rm v >$. The top panel shows the resulting posterior for $\sigma_{\rm MCMC}$ and the bottom right panel the posterior for $<\rm v>$.
\end{itemize}

\section{Errors in individual velocity measurements}
\label{app:errors}

\begin{figure}
    \centering
   \includegraphics[scale = 0.62]{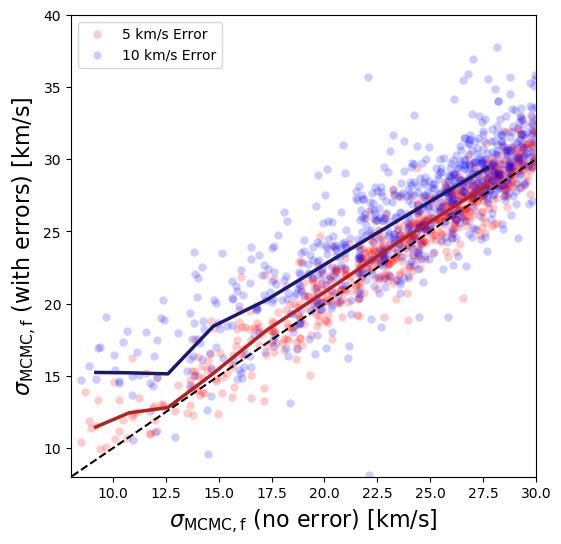}
    \caption{Impact of adding errors to individual GC's velocity measures on the recovered velocity dispersion of the system using the MCMC method with a flat prior. The horizontal axis shows $\sigma_{\rm MCMC,f}$ (assuming no errors) and the y-axis shows for the same GC systems the $\sigma_{\rm MCMC,f}$ calculated assigning to each GC velocity an error drawn from a Gaussian distribution with dispersion $5$ (red) and $10$ (blue) km s$^{-1}$. Solid lines with the same color indicate the median at fixed $\sigma_{\rm MCMC,f}$.}
    \label{fig:mcmc_err}
\end{figure}

We explore in Fig.~\ref{fig:mcmc_err} the effect of adding measurement errors to the individual velocity of GCs in each galaxy. We compare the MCMC (flat prior) velocity dispersion calculated with and without errors, where errors have been modeled assuming a Gaussian distribution of 5 and 10 km s$^{-1}$ dispersion (red and blue, respectively). These values have been chosen to coincide with typical velocity errors in recent observations of dwarf galaxies \citep{Toloba2016, Toloba2018, vanDokkum2018b}. 

One can expect that these added uncertainty will only be relevant in objects where the intrinsic velocity dispersion of the GC system is of the order of the added errors to each individual GC velocity. We therefore show in Fig.~\ref{fig:mcmc_err} a subsample of our galaxies with intrinsic $\sigma_{\rm MCMC,f} \leq 30$ km s$^{-1}$. As expected, we find that the addition of errors will tend (on the median) to increase the velocity dispersion estimates, with an impact that naturally depends on the level of errors included. 

%%%%%%%%%%%%%%%%%%%%%%%%%%%%%%%%%%%%%%%%%%%%%%%%%%%
%%%%%%%%%%%%%%%%%%%%%%%%%%%%%%%%%%%%%%%%%%%%%%%%%%%

The maximum effect is found for our lowest velocity dispersion objects, where the overestimation on the median may reach $20\%$ in the case of 10 km s$^{-1}$ errors. Note that this quickly decreases to 5\% if the errors are instead 5 km s$^{-1}$. The solid red and blue lines indicating the median MCMC determination including errors show that the systematic overestimation decreases as the intrinsic velocity dispersion increases, being negligible for objects with $\sigma_{\rm MCMC,f} \sim 25$ km s$^{-1}$ and above. This study indicates that the inclusion of observational errors in our calculations does not qualitatively change the results and conclusions presented in our paper.

\section{Impact of non-Gaussian distributions on Confidence Intervals}
\label{app:gaussian}

%%%%%%%%%%%%%%%%%%%%%%%%%%%%%%%%%%%%%%%%%%%%%%%%%%%
\begin{figure}
    \centering
   \includegraphics[width=\columnwidth]{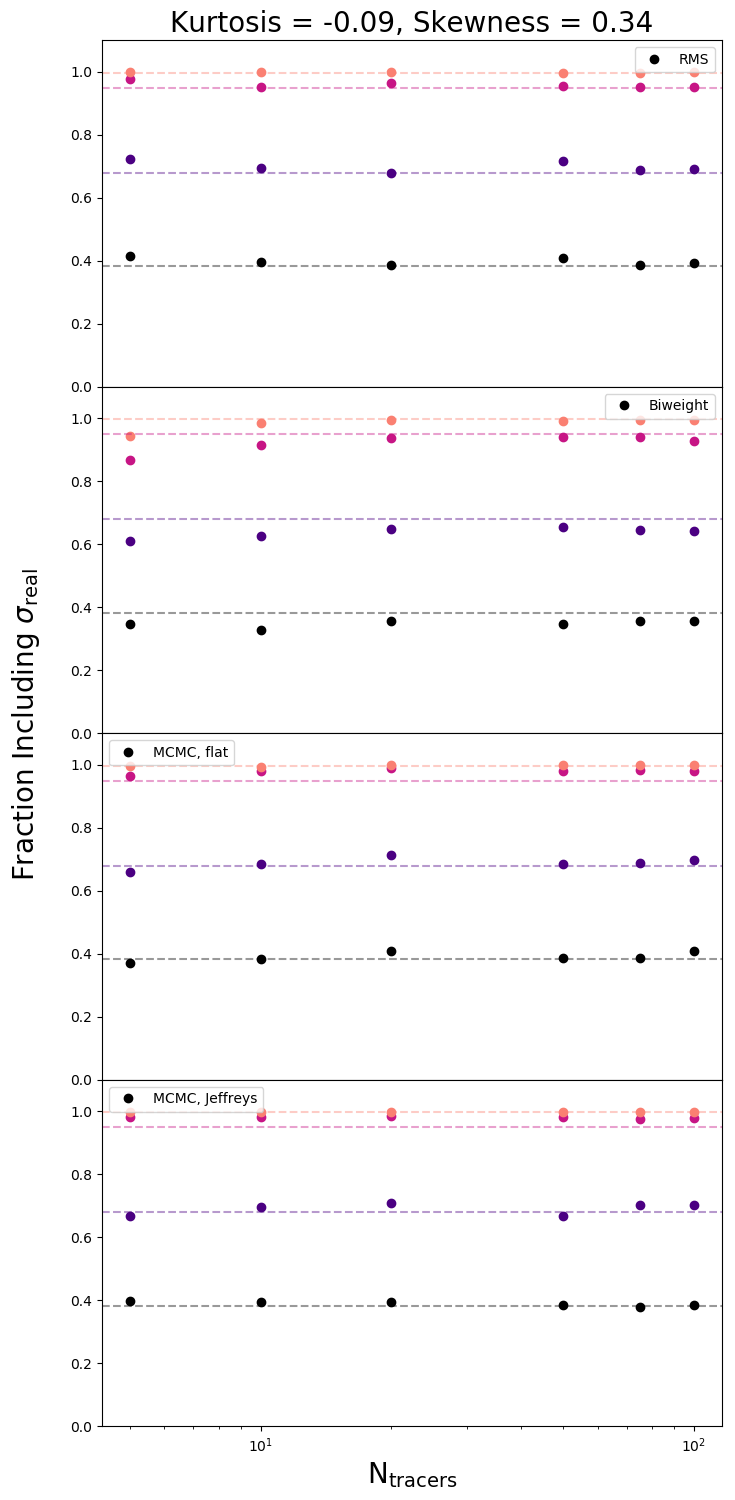}
	\caption{For 1000 realizations of 5, 10, 20, 50, 75, and 100 GCs, the fraction of times that $\sigma_{\rm real}$ of the underlying distribution of GC candidates falls within the specified confidence interval (38.3$\%$ (black), 68$\%$ (purple), 95$\%$ (magenta) and 99.7$\%$ (orange)) of the specified $\sigma$ estimate for the realization. Trials from top to bottom are for $\sigma_{\rm rms}$, $\sigma_{\rm biweight}$, $\sigma_{\rm MCMC,f}$ and $\sigma_{\rm MCMC.j}$. This particular subhalo, from the top panel of Fig.~\ref{fig:introfig}, has a relatively Gaussian distribution of GC candidates. The Gaussian confidence seem to hold quite well across all $N_{\rm GC}$ with the exception for $\sigma_{\rm biweight}$ for which the confidence intervals are underestimated. }
    \label{fig:frac_gauss}
\end{figure}
%%%%%%%%%%%%%%%%%%%%%%%%%%%%%%%%%%%%%%%%%%%%%%%%%%%

Confidence intervals represent the probability (or fraction of times) that the true variance $s^2$ of a sample with $N_{\rm tot}$ events falls within the variance $d^2$ plus/minus the confidence interval of a given subsample with $N$ objects (where $N < N_{\rm tot}$). This confidence intervals have a well-known functional form in the case of an underlying Gaussian distribution, an assumption commonly made to estimate the accuracy of velocity measurements in observations. In this Appendix we test how well the Gaussian confidence intervals perform for $5$ individual objects in our sample when using each of the three methods to measure velocity dispersion explored in this paper: r.m.s, biweight and MCMC. 

We start by using the galaxy introduced in the upper row of Fig.~\ref{fig:introfig}, which shows a nearly Gaussian line-of-sight velocity distribution (see upper right panel in the same figure). The kurtosis and skewness for the GCs candidates in this object are $-0.09$ and $0.34$, respectively. We sample $1000$ times $N=5$, $10$, $20$, $50$, $75$ and $100$ GCs out of the $\sim 400$ GC candidates that remain bound to this galaxy at $z=0$. For each set of samplings, we count the fraction of times than the true variance of all {\it candidate} GCs is contained within the variance of each random sampling with $N$ tracer GCs plus the confidence interval computed assuming a Gaussian distribution. 

The upper panel in Fig.~\ref{fig:frac_gauss} shows the result of such exercise as a function of the number of tracers $N$ selected. We show with circles the results for confidence interval levels: $38.3\%$ black, $68\%$ (purple), $95\%$ (magenta) and $99.7\%$ (orange). Dashed horizontal lines highlight the position of each level in the plot. For a perfectly Gaussian distribution and a reliable method to estimate velocity dispersion, one would expect the symbols to follow these horizontal lines. 

%%%%%%%%%%%%%%%%%%%%%%%%%%%%%%%%%%%%%%%%%%%%%%%%%%%
\begin{figure*}
    \centering
   \includegraphics[width=0.48\columnwidth]{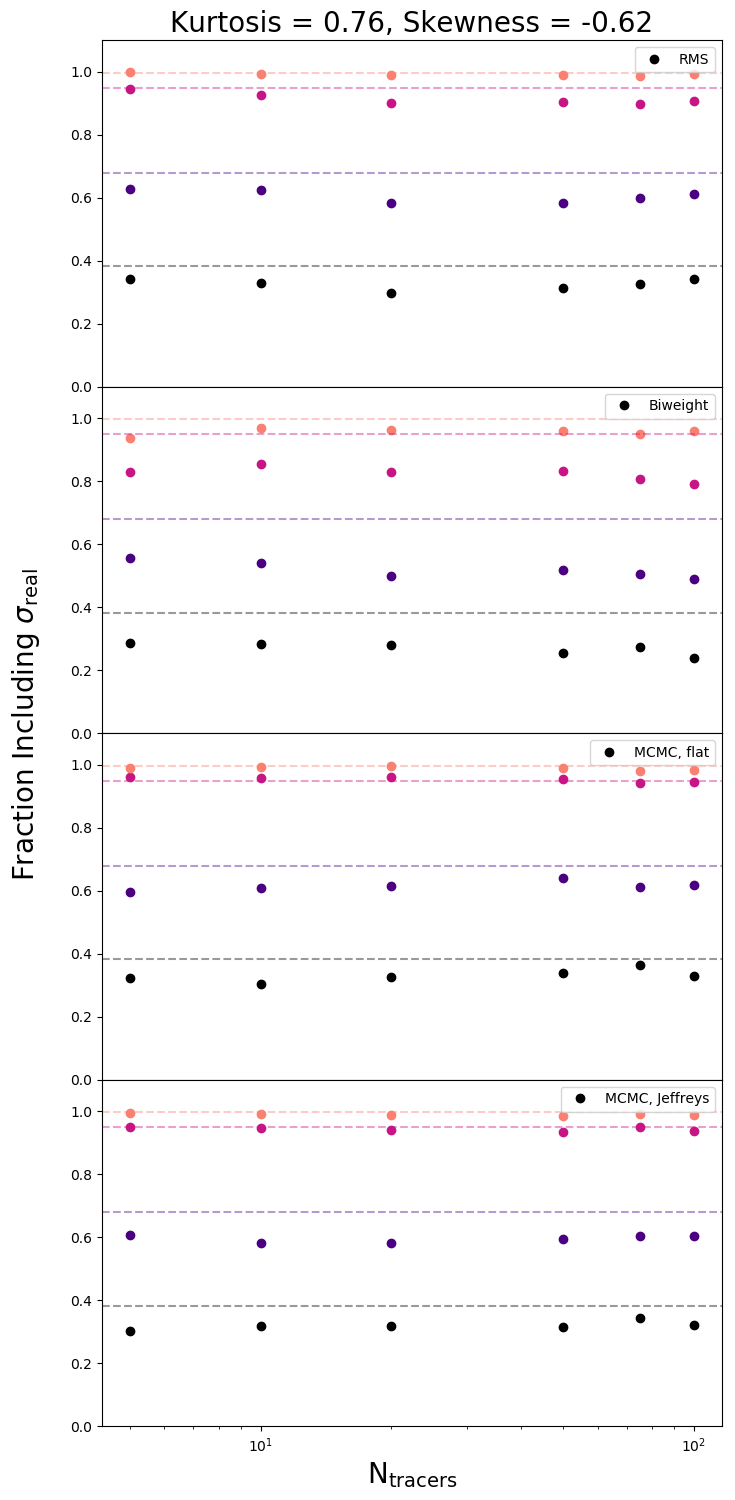}
   \includegraphics[width=0.48\columnwidth]{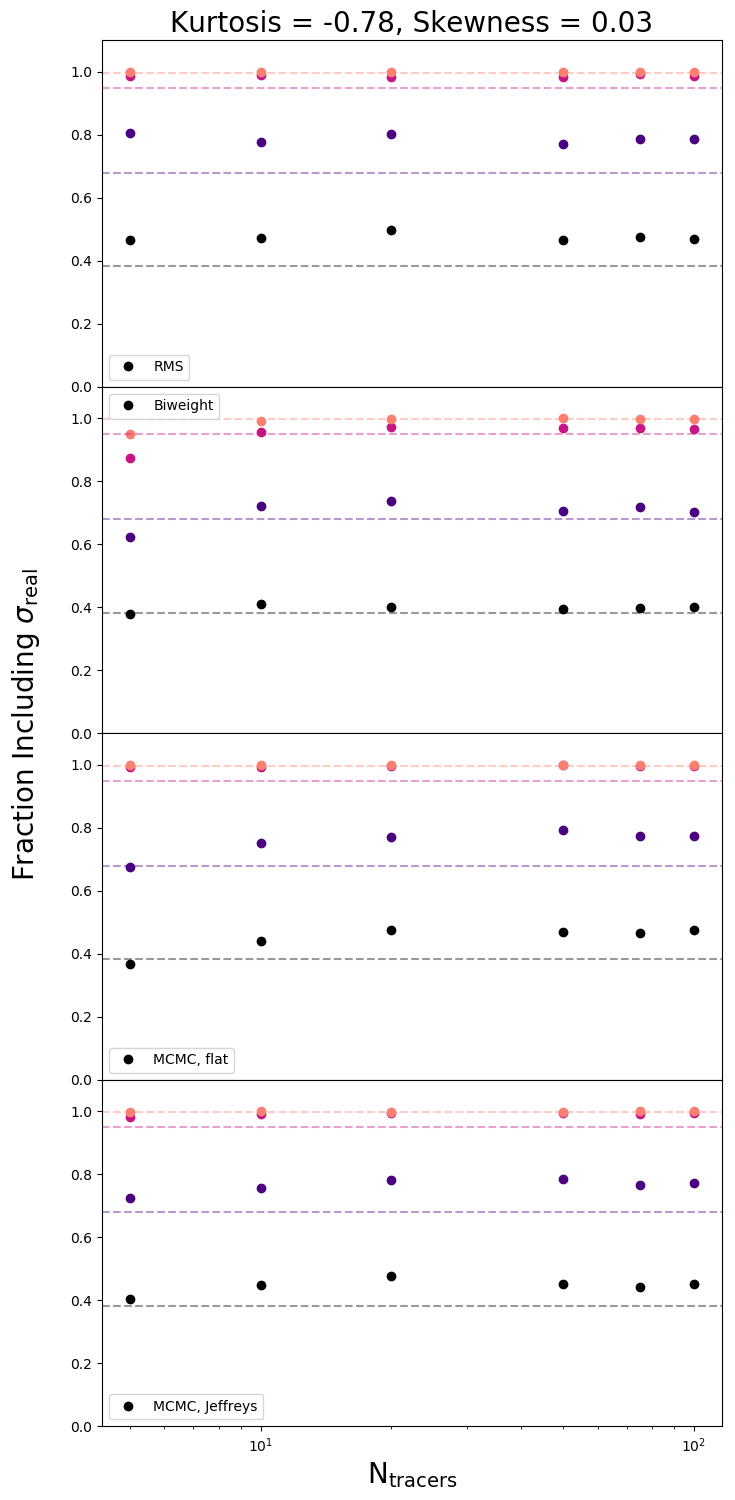}
   \includegraphics[width=0.48\columnwidth]{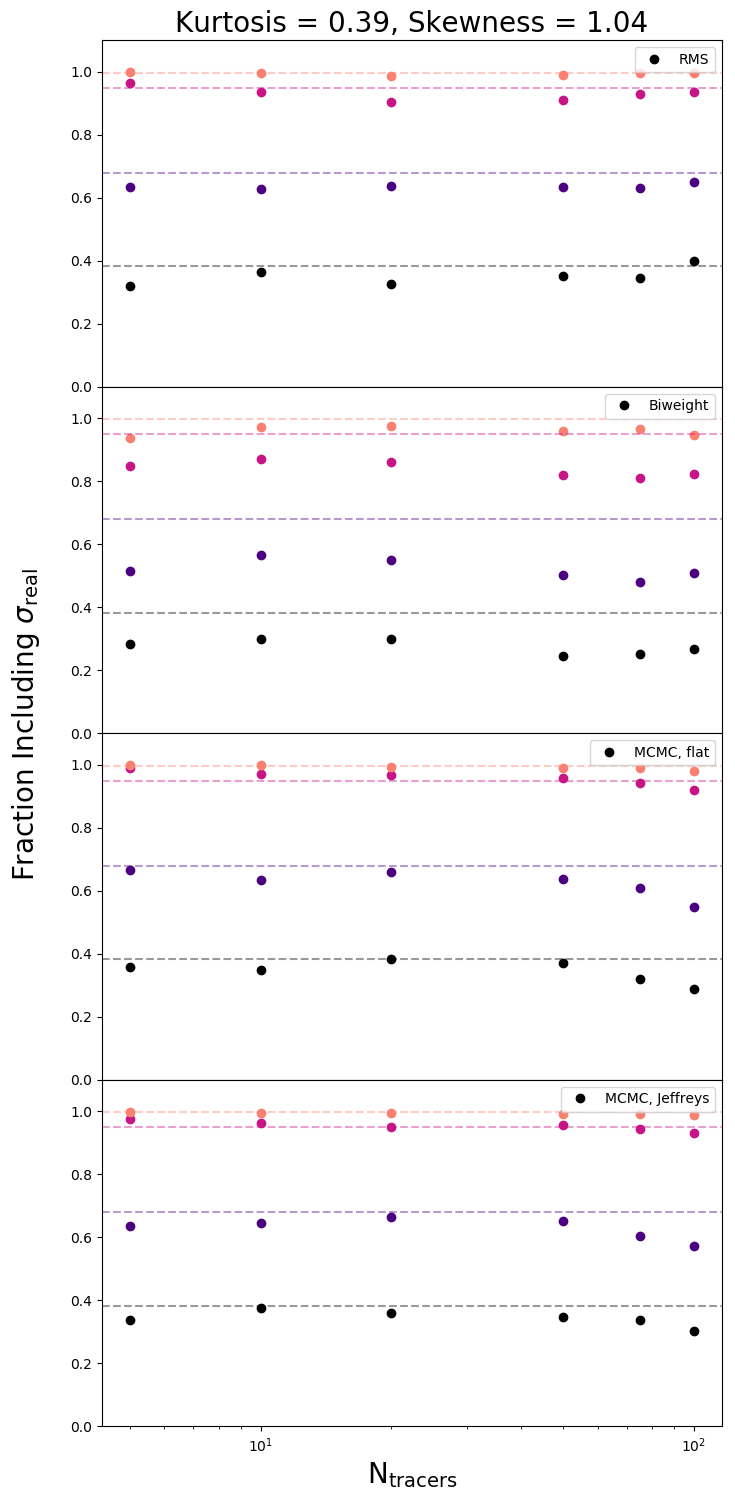}
   \includegraphics[width=0.48\columnwidth]{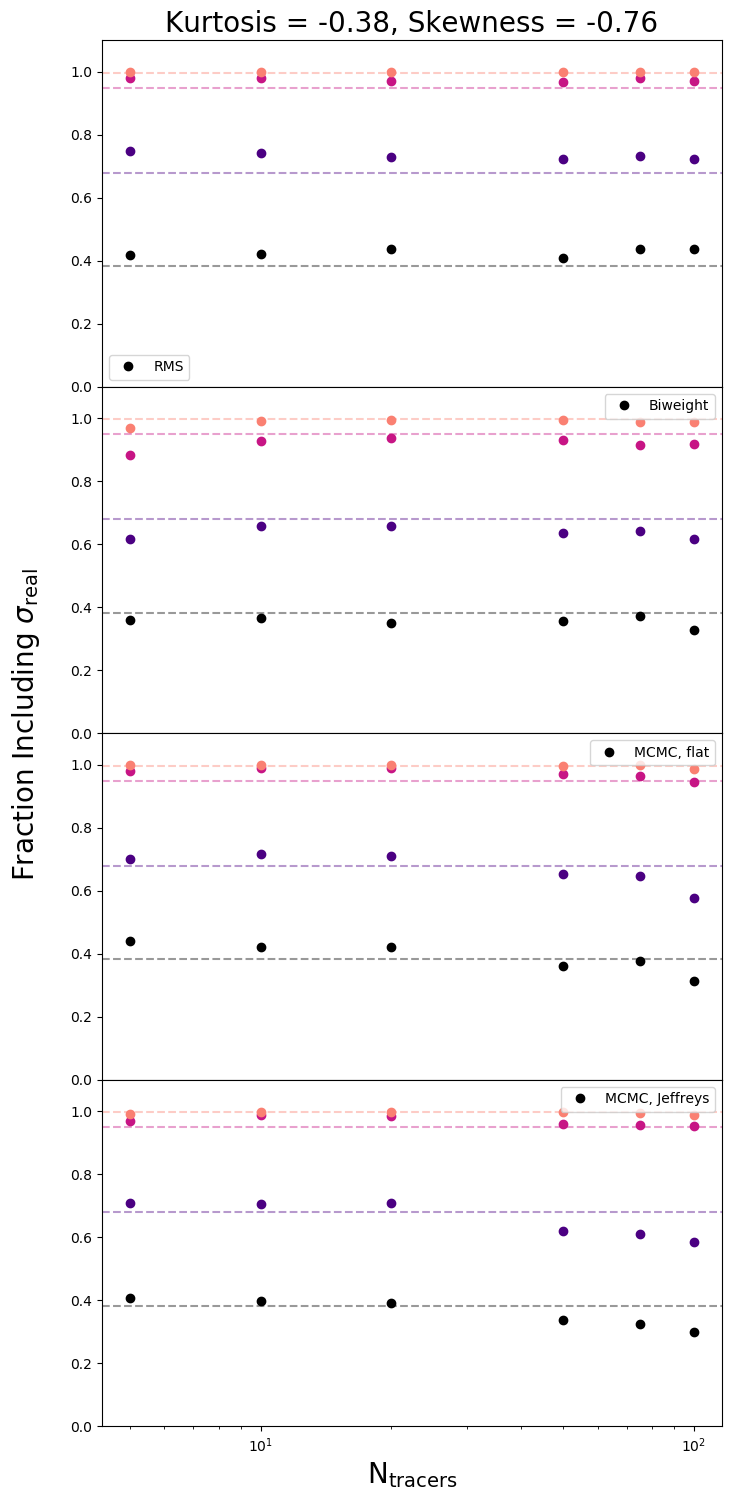}
	\caption{For 1000 realizations of $N_{\rm tracers}=$ 5, 10, 20, 50, 75, and 100 GCs, the fraction of times that $\sigma_{\rm real}$ of the underlying distribution of GC candidates falls within the corresponding confidence interval (38.3$\%$ (black), 68$\%$ (purple), 95$\%$ (magenta) and 99.7$\%$ (orange)) for the different $\sigma$ estimates. Trials from top to bottom in each panel are for $\sigma_{\rm rms}$, $\sigma_{\rm biweight}$, $\sigma_{\rm MCMC,f}$ and $\sigma_{\rm MCMC,j}$. From left to right, each panel shows the effect of large positive kurtosis, large negative kurtosis, large positive skewness, and large negative skewness on the correctness of Gaussian confidence intervals. All methods seem to suggest that confidence intervals are underestimated in the case of negative kurtosis (points above horizontal lines) while the opposite is true for positive kurtosis. The trend with number of tracers is rather weak.}
    \label{fig:frac_skew}
\end{figure*}
%%%%%%%%%%%%%%%%%%%%%%%%%%%%%%%%%%%%%%%%%%%%%%%%%%%

We find that this is the case of the r.m.s velocity dispersion estimate in this galaxy roughly independent of the number of tracers (upper panel in Fig.~\ref{fig:frac_gauss}). Similarly, computing the velocity dispersion using MCMC (either with flat or Jeffreys prior, bottom 2 panels in Fig.~\ref{fig:frac_gauss}) yields a similar result. In this case, the confidence interval is not computed from the Gaussian form, but extracted directly from the PDFs of the MCMC method. 

In the case of $\sigma_{\rm biweight}$ (top second panel in Fig.~\ref{fig:frac_gauss}), assuming Gaussian confidence intervals seems to slightly overestimate the accuracy (dashed lines are above the calculated symbols), specially for a number of tracers 10 and below. However, the effect is only mild. 

We repeat this calculation using $4$ galaxies that deviate more substantially in either kurtosis or skewness from a Gaussian distribution (those highlighted in Fig.~\ref{fig:tidalstrip}). We show this in Fig.~\ref{fig:frac_skew}. We find that all methods show in general similar trends: high positive kurtosis results on underestimated confidence intervals (symbols below the dashed lines) while high negative kurtosis means that measurements are more accurate than expected from a perfectly Gaussian distribution (symbols above corresponding dashed lines). Similar trends might be found for deviations in skewness (rightmost two panels in Fig.~\ref{fig:frac_skew}), although the impact seems smaller than in the case of kurtosis. The dependence with the number of tracers is rather weak. 

Such exercise (and the ratio between the symbols and the horizontal levels) can now be applied to the whole sample to derive, for each galaxy, a correction to the confidence intervals calculated assuming a Gaussian distribution. This is shown in Fig.~\ref{fig:error_correc} of the main text.

\bsp	% typesetting comment
\label{lastpage}
\end{document}